\documentstyle[12pt,epsf]{article}
\begin{document}

    \title{ Neutrino Oscillations in Noisy Media}
    \author{F. N. Loreti and A. B. Balantekin\\
     Department of Physics, University of Wisconsin-Madison,\\
     Madison, WI 53706\\
     and\\
     Institute for Nuclear Theory, University of Washington,\\
     Seattle, WA 98195}
     \maketitle

     \begin{abstract}
   We develop the Redfield equation for delta-correlated gaussian noise and
apply it to the case of two neutrino flavor or spin precession in the
presence of a noisy matter density or magnetic field, respectively.
The criteria under which physical fluctuations can be well
approximated by the delta-correlated gaussian noise for the above cases
are examined. Current limits on the possible neutrino magnetic moment
and solar magnetic field suggest that a reasonably noisy solar magnetic
field would not appreciably affect the solar electron neutrino flux.
However, if the solar electron density has fluctuations of a few percent
of the local density and a small enough correlation length,
the MSW effect is suppressed for a range of parameters.
       \end{abstract}

\section{INTRODUCTION}

\indent

Neutrino oscillations in the presence of matter and magnetic fields have been
an area of intense study for approximately the last ten years. In the
Mikheyev - Smirnov - Wolfenstein (MSW)
effect, electron neutrinos on their journey from the core, are resonantly
transformed into muon or tau neutrinos \cite{msw,us}.
If neutrinos are Majorana fermions
with transition magnetic moments, they can undergo a magnetic resonant
transformation into muon or tau antineutrinos \cite{lam}.
Neutrinos from a supernovae
explosion can be transformed from one flavor to another as they pass
through the outer part of the star \cite{pant}.
In many stellar situations, the
matter density and/or magnetic fields may fluctuate about a mean value.
Some well known
examples where fluctuations are likely to exist include the magnetic field
and the matter density in the solar
convective zone and also the turbulence of the post-supernovae matter which has
been blown off by the explosion.  A general approach to the neutrino
oscillations in inhomogeneous matter was developed in Ref. \cite{sawyer}.
A Study of matter fluctuations which are not random,
but harmonic \cite{koonin,wick}, or occur as a jump-like change in the solar
density \cite{denvar}, are available in the literature.
Matter currents and density changes effect neutrino
flavor oscillations in a similar way and have also been
examined  in Ref. \cite{wick}. Although matter current effects become
important only if the velocity is somewhat close to the speed of light, noisy
mixing of matter would also mimic a fluctuating matter density.
A priori,
fluctuations in such fields may be well approximated by random noise added to
an average value. In this paper we show how such noise will affect neutrino
oscillations for the situation in which the correlation length of the randomly
fluctuating part of either the matter density or the magnetic field is small
compared with the neutrino oscillation length.

The case of neutrino spin precession in a noisy magnetic field was considered
by Nicolaidis for neutrinos in vacuum \cite{nic}.
The noise was taken to be well
approximated by a delta-correlated gaussian distribution with the result
that the normal oscillations become damped with a relaxation time of
$t_{rel} = (2\mu^2 <B_r^2> \tau_c)^{-1}$, where $B_r$ is the randomly
fluctuating part of the magnetic field and $\tau_c$ is its correlation
time. Enqvist and Semikoz considered neutrino oscillations in randomly
fluctuating magnetic field, approximately, by averaging the coefficients
in the third order differential equation for the z-component of the
neutrino spin, in a constant matter density \cite{enq}.
Their result suggested that
the effect of the randomly fluctuating part of the magnetic field
was similar to the effect of a larger (constant) matter density, hence
reducing the neutrino precession.
Our result is quite different; namely that both a noisy magnetic field
and a noisy density (for a constant averaged density) act
to depolarize the neutrinos. For a noisy magnetic field, if the probability
of transition is greater than one half without the random fluctuations, the
inclusion of the random fluctuations will reduce the transition probability,
and if the transition probability without the random fluctuations is less than
one half, the inclusion of the fluctuations will increase the transition
probability. If the randomly fluctuating part of the magnetic field is
strong enough or is allowed to act for a long enough time, a complete
depolarization of particles-antiparticles will occur.
For a noisy density, we find that the MSW transition probability
is suppressed. For the case of strongly adiabatic MSW transitions and
large fluctuations, the averaged transition probability saturates at one
half.

In Section II, we develop the equations governing the time evolution of
the averaged
probabilities of being found in the N${th}$ level of an N-level system,
subject to random and non-random potentials.
These equations are derived for the case where
the randomly fluctuating part of the field is taken to be a delta-correlated
gaussian. In Section III, we show several analytically solvable examples for
the
case of a two level problem, and numerically examine cases the
analytical solutions of which are not instructive. In Section IV, we
investigate
the conditions under which a real, physical fluctuating field will be well
approximated by the equations developed for the delta-correlated gaussian
case. We then apply these conditions and examine the cases of spin-flavor
and flavor precession of neutrinos in the sun. Section V presents
a discussion of results and our conclusions.

\section{FORMULATION OF THE APPROACH}

\indent

We first consider the general case of
an N-level system with time dependent level splitting and transition
terms, in the presence of an additional randomly fluctuating term. Since
we are interested in the ensemble average of the probability of finding the
system in a given level at time t, we consider the density operator defined
as,
\begin{equation}
\label{eq1}
\hat {\rho} \equiv { \psi \otimes \psi^{\dagger}},
\end{equation}
where,
\begin{equation}
\label{eq2}
\psi(t) = \left(
\begin{array}{c}
\psi_1(t) \\
\psi_2(t) \\
\vdots \\
\psi_N(t)
\end{array}\right).
\end{equation}
Assuming that ${ \psi}$ obeys the Schr\"odinger-like equation,
\begin{equation}
\label{eq3}
i {d\over dt} \psi = \hat H \psi,
\end{equation}
$\hat {\rho}$ obeys the equation,
\begin{equation}
\label{eq4}
i{d\over dt}\hat {\rho} = [\hat H,\hat {\rho}];
\end{equation}
where the Hamiltonian is taken to be linear in fluctuating field, i.e.
\begin{equation}
\label{eq5}
\hat H =\hat H_0(t) + B(t)\hat M^{\prime},
\end{equation}
where $ \hat M^{\prime}$ is independent of time. Eq. (4) becomes,
\begin{equation}
\label{eq6}
i{d\over dt}\hat {\rho}_I = [\hat H_I,\hat {\rho}_I],
\end{equation}
where,
\begin{eqnarray}
\label{eq7}
\hat {\rho}_I &=& \hat U_0^{\dagger} \ \hat {\rho} \ \hat U_0, \nonumber \\
\hat M &=& \hat U_0^{\dagger} \ \hat M^{\prime} \ \hat U_0, \\
\hat H_I &=& B(t)\hat M, \nonumber
\end{eqnarray}
and $\hat U_0$ satisfies,
\begin{equation}
\label{eq8}
i{d\over dt}\hat U_0 = \hat H_0 \hat U_0.
\end{equation}
This equation can be solved by iteration as,
\begin{eqnarray}
\label{eq9}
\hat {\rho}(t) &=& \hat {\rho}_0 - i \int_0^tdt_1B(t_1)[\hat
M(t_1),\hat {\rho}_0] \nonumber \\
&-& \int_0^t\int_0^{t_1}dt_1 dt_2 B(t_1)B(t_2)\bigl[\hat M(t_1),[\hat
M(t_2),\hat {\rho}_0]
\bigr]
+ \cdots
\end{eqnarray}
We now assume that B(t) is
such that,
\begin{equation}
\label{eq10}
<B(t)>=0, \quad <B(t_1)B(t_2)>=\alpha^2 f(\mid t_2 - t_1 \mid),
\end{equation}
with the average of all higher odd products of B vanishing, and all higher
even products given by the sum of all possible products of f's for example,
\begin{equation}
\label{eq11}
<B(t_1)B(t_2)B(t_3)B(t_4)>= \alpha^4 [f_{12}f_{34} + f_{13}f_{24} +
f_{14}f_{32}],
\end{equation}
where, $f_{12} \equiv f(\mid t_2 - t_1 \mid)$.
This will result in only
the even products contributing to $<\hat {\rho}(t)>$.

Further analytic progress can be made if the random potential is a
delta correlated gaussian distribution, where $f(\mid t_1 - t_2 \mid)$
becomes $2 \tau \delta (t_1 - t_2)$. Averaged values of functions of the
random field B(t) can then be expressed as a path integral,
\begin{equation}
\label{eq12}
<g(B(t))> = \int_{-\infty}^{\infty} {\cal D}[B(t)] g(B) e^{ -\int_0^t
dt {{B^2} /2k}},
\end{equation}
where,
\begin{equation}
\label{eq13}
{\cal D} [B(t)] = \prod_i dB(t_i) \sqrt{{\Delta t}\over {2k \pi}},
\end{equation}
and
$k=2 \alpha^2 \tau$. This reproduces the form of the averages of the even power
products, and the integrals in Eq. (9) can be explicitly calculated with the
result,
\begin{eqnarray}
\label{eq14}
<\hat {\rho}_I(T)> = \hat {\rho}_0 &-& \alpha^2 \tau \int_0^T dt_1
\bigl[\hat M(t_1),[\hat M(t_1),\hat {\rho}_0]\bigr]
\phantom{ooiueoiueoiueoiueoiue}
\nonumber \\
&+& \alpha^4 \tau^2 \int_0^T dt_1
[\hat M(t_1),[\hat M(t_1),\int_0^{t_1}dt_2 [\hat M(t_2),[\hat
M(t_2),\hat {\rho}_0]]]] \\
&-& \cdots \nonumber
\end{eqnarray}
This, however, is just the iterative expansion of the differential
equation,
\begin{equation}
\label{eq15}
{d\over dt}<\hat {\rho}_I(t)> = -\alpha^2 \tau \bigl[\hat M(t),[\hat
M(t),<\hat {\rho}_I(t)>]
\bigr].
\end{equation}
One may equally well express this result as,
\begin{equation}
\label{eq16}
{d\over dt}<\hat {\rho}(t)> = -\alpha^2 \tau \bigl[\hat M^{\prime},[\hat
M^{\prime},
<\hat {\rho}(t)>]\bigr] -i[{ \hat H_0}(t),<\hat {\rho}(t)>].
\end{equation}
Eq. (16), in general, defines a set of N$^2$ coupled first order
linear differential equations with the constraint that Tr$\hat {\rho} =
1$.  This is a matrix form of the Redfield equation \cite{kam}.
In the case of neutrinos moving in a varying background, Eq. (16) is
equivalent to Eq. (49) in Ref. \cite{sawyer}.

\section{Analytic Examples}

\indent

Perhaps the simplest examples are those for which one could obtain an
integral expression for the
exact solution for the probabilities of the system being found in the
$N^{th}$ level at time t, and then perform the averaging using Eq. (12).
Such an expression is possible, in general,
only in the case where all N levels are degenerate. For example,
consider the case in which the Hamiltonian is an $N \times N$ matrix which is
of
the form $\hat H(t) = (B_0(t) + B_r(t)) \times \hat M$, where $\hat M$ has
zeros
on the diagonal and ones everywhere else. One could, for this case, obtain
an integral expression for the probabilities and then average them.
Instead, we take $\hat H_0(t)= B_0(t)\hat M$
giving $\hat U_0(t) = {\rm exp}( \int_0^t dt B_0(t) \hat M)$ and use Eq. (15)
which
becomes,
\begin{equation}
\label{eq17}
{d\over dt}<\hat {\rho}_I(t)> = -{k(t)\over 2} \bigl[\hat M,[\hat
M,<\hat {\rho}_I(t)>]\bigr],
\end{equation}
where
\begin{equation}
\label{17a}
k(t)/2 = \alpha^2 (t) \tau .
\end{equation}
Taking the commutator of $\hat M$ with
${d\over dt}<\hat {\rho}_I(t)>$, and noting that $\hat M^2 = (N-1) +
(N-2)\hat M$, one obtains,
\begin{equation}
\label{eq18}
{d\over dt}[\hat M, <\hat {\rho}_I(t)>] = -{N^2\over 2} k(t) [\hat M,
<\hat {\rho}_I(t)>].
\end{equation}
This equation is easily solved with the solution,
\begin{equation}
\label{eq19}
<\hat {\rho}(T)> = { \hat X(T)} - N^{-2}(1 - e^{-{N^2\over2} \int_0^T dt k(t)})
[\hat M,[\hat M,{ \hat X(T)}]]
\end{equation}
where ${ \hat X(t) = \hat U_0(t) \hat {\rho}(0) \hat
U_0^{\dagger}(t)}$. This reproduces
the result of Nicolaidis for the case where $\hat M^{\prime}$ is the Pauli
matrix
$\sigma_x$, and the initial state is $\psi_1(0)=1, \psi_2(0)=0$.

A second analytically solvable example is a two level system, with an arbitrary
time dependent level splitting, in a purely random field. This would
correspond to the case of matter-enhanced neutrino spin precession in a noisy
magnetic field.
In this case, there
exists no analytic expression for the probabilities which one could in
principle average. We begin by using Eq. (16) with $\hat H_0 = A(t)\sigma_z +
B_0(t)\sigma_x$ and $\hat M^{\prime} = \sigma_x$.
Defining $ r(t) \equiv {1 \over 2} (<\hat {\rho}_{11}> -
<\hat {\rho}_{22}>), x = 2 Re(<\hat {\rho}_{12}>)$ and $y \equiv 2
Im(<\hat {\rho}_{12}>)$, one obtains,
\begin{equation}
\label{eq20}
{d\over dt}\left(
\begin{array}{c}
r \\ x \\ y
\end{array}\right) = -2 \left(
\begin{array}{ccc}
k & 0 & B_0 \\ 0 & 0 & A \\ -B_0 & -A & k
\end{array}\right)
\left(\begin{array}{c}
r \\ x \\ y \end{array}\right).
\end{equation}
If $B_0 = 0$ Eq. (21) is trivially solved giving,
\begin{equation}
\label{eq21}
r(T)=r(0)e^{-2\int_0^Tdt k(t)},
\end{equation}
resulting in,
\begin{equation}
\label{eq22}
<|\psi_1(T)|^2> = {1 \over 2} (1 -  e^{-2\int_0^Tdt k(t)}) +
|\psi_1(0)|^2 e^{-2\int_0^Tdt k(t)}.
\end{equation}
This result is surprising in that it is independent of the level splitting and
identical to the result for zero level splitting.
If $B_0 \neq 0$, one
could still obtain an analytic expression for the averaged probabilities
if the level splitting is constant, although such an expression is not
instructive due to the complicated nature of third roots. We instead
calculate numerically cases of constant level splitting. Figure (1a) through
(1d) show the probability $|\psi_1 (t)|^2$ for $A=0, A=0.5, A=2$ and $A=10$
respectively, for $B_0=1$ and $k=0.05$. One observes from these figures that
as $A$ is increased, the oscillations become heavily damped while the
exponentially decreasing upper envelope remains essentially unchanged. This
qualitatively indicates that even when one is far from resonance ($A=0$ for
the case in which A is time dependent),
one can still obtain complete depolarization for such a randomly
fluctuating field.

Another possibility occurs if one has the case of a noisy diagonal term
with a non-noisy off-diagonal term, namely $H= (A_0(t) + A_r(t))\sigma_z +
B_0(t)
\sigma_x$. This case would correspond to a noisy matter density in the
context of matter-enhanced neutrino oscillations.
Eq. (16) becomes,
\begin{equation}
\label{eq23}
     {d\over dt}\left(
     \begin{array}{c}
         r \\ x \\ y
     \end{array}\right) = -2 \left(
     \begin{array}{ccc}
         0 & 0 & B_0(t) \\ 0 & k & -A_0(t) \\ -B_0(t) & A_0(t) & k
      \end{array}\right)
      \left(\begin{array}{c}
         r \\ x \\ y
      \end{array}\right).
\end{equation}
If $A_0(t)=0$ and $B_0$ is time independent, one can obtain an exact solution
which,
for the initial condition $\psi_1(0)=1, \psi_2(0)=0$ and $4B_0^2 > k^2$ is,
\begin{equation}
\label{eq24}
|\psi_1(t)|^2 = {1 \over 2} + {1 \over 2} e^{-kt}\big[ {{k}\over
\omega}\sin \omega t
+ \cos\omega t \big],
\end{equation}
where $\omega = \sqrt{4B_0^2 - k^2}$.
We again numerically calculate
examples in which the non-random diagonal term is  non-zero. Figures (2a)
through (2d) show $|\psi_1(t)|^2$ for $A_0=0, A_0=0.5, A_0=2$ and
$A_0=10$ respectively
for $B_0=1$ and $k=0.05$. In in contrast to the case of a noisy off-diagonal
term, in this case
one observes that as the oscillations
are damped, the exponential relaxation time increases. Thus, if one is far
from resonance, the depolarization is very much suppressed, as is physically
reasonable.

\section{NEUTRINOS IN NOISY SOLAR FIELDS}

\indent

In Appendix I, we detail the expansion of the density matrix in the case
where the randomly fluctuating field has a finite correlation time, with
the result that Eq. (16) can accurately describe a real physical
situation if,
\begin{equation}
\label{eqn25}
( d/dt (\log \alpha (t)) + [\hat H_0(t),\hat M^{\prime}]) \tau_c \ll 1,
\end{equation}
for all t of interest, where $\tau_c$ is the correlation time of the
randomly fluctuating field and $\alpha$ is it's root mean square
value, which may be time dependent. If this is satisfied for a given problem,
$k(t)$ is given by,
\begin{equation}
\label{eqn26}
k(t) = \beta \alpha^2 \tau_c(t),
\end{equation}
where $\beta$ is a factor
of order unity, equal to $2$ for our choice in Appendix I of a theta
function correlation.

\subsection{Spin Precession in a noisy solar magnetic field}

\indent

We first consider the case of matter-enhanced
spin-flavor precession of Majorana neutrinos
with a negligibly small MSW mixing angle. In this case the evolution
equation is two by two, and is given by \cite{lam},
\begin{equation}
\label{eqn27}
     i{d\over dr}\left(
        \begin{array}{c}
            \psi_{\nu_e} \\ \psi_{\overline{\nu}_{\mu}}
        \end{array}\right) = \left(
        \begin{array}{cc}
           -{{\Delta m^2} \over 4E} + a_e & \mu (B(r)+ B_r(r)) \\
            \mu (B(r) + B_r(t))& {{\Delta m^2} \over 4E} -a_{\mu}
        \end{array}\right)\left(
        \begin{array}{c}
             \psi_{\nu_e} \\ \psi_{\overline{\nu}_{\mu}}
        \end{array}\right),
\end{equation}
where $\Delta m^2 \equiv m_2^2 - m_1^2$ with $m_i$ the masses of the mass
eigenstate neutrinos, E is the neutrino
energy, $\mu$ is the neutrino transition magnetic moment, $B$ and $B_r$ are
the magnetic field and its noise, and $a_e$ and $a_{\mu}$ are the matter
potentials given by,
\begin{equation}
\label{eqn28}
a_e = {1\over \sqrt{2}}G_F(2N_e - N_n), \qquad a_{\mu}= {-1\over \sqrt{2}}G_F
N_n,
\end{equation}
where $N_e(r)$ and $N_n$(r) are the electron and neutron number densities.

{}From the discussion in the previous section of a noisy off-diagonal
term, we can estimate whether a reasonably noisy magnetic field
can effect the transition probability. In order for the random fluctuations
to have much effect, $2k\Delta r \sim 1$. The condition
for this real process to be approximated by a delta-correlated gaussian,
namely Eq. (26), results, for a constant rms value of the
the magnetic field, in the condition,
\begin{equation}
\label{eqn29}
\tau_c \times[-{{\Delta m^2} \over 2E} +
{\sqrt{2}}G_F(N_e - N_n)] \ll 1.
\end{equation}

Perhaps the most likely place a noisy magnetic field would exist is
in the solar convective zone which extends from approximately
$0.7R_{\odot}$ to the surface \cite{bac}. The magnitude of the
solar magnetic field observed at the surface of the sun can reach local values
of several kilogauss in magnetic storms, and it is thought that it may
reach values of 100 kG near the bottom of the convective zone \cite{sof}.
While a large
( $\sim 100$kG ) field which
extends through out a large fraction of the convective zone is
not ruled out, such a large, extensive field is thought to be unlikely
\cite{schram}.
Field strengths as large as several times $100$kG could exist if limited
to extensions of about $10-100$ km.
In order to estimate the maximum effect of magnetic field fluctuations,
we will assume
that the solar magnetic field from $0.7R_{\odot}$ to $0.85R_{\odot}$ can
be considered to be randomly fluctuating with zero mean and a rms value
of $100$kG. We take the neutrino transition magnetic moment to be
$\mu = 3 \times 10^{-12}\mu_{B}$, in accord with the maximum bound from
plasmon decay in pre-helium flash red giant stars \cite{raf}. By choosing a
value of
$\Delta m^2 / 2E$, one can determine the maximum $\tau_c$ (assumed to be
constant) such that
Eq. (30) is satisfied throughout the above region in the solar convective zone.
Therefore, we take $\tau_c$
to given by,
\begin{equation}
\label{eqn30}
\tau_c = 0.1 \times (|-{{\Delta m^2} \over 2E} +
{\sqrt{2}}G_F(N_e - N_n)|_{max})^{-1}.
\end{equation}
Given that Eq. (30) is obeyed, the average electron neutrino probability
after traversing this region is given by Eq. (23), with $\psi_e(0) = 1$,
and for small $2k\Delta r$ the muon antineutrino average probability
is approximately given by $k \Delta r$.
In Figure (3) we plot
the value of $k\Delta r = 4<(\mu B)^2>\tau_c (0.15R_{\odot})$ as a function
of $\Delta m^2 / E$. One observes that the maximum occurs at
about $1\times 10^{-8} {\rm eV^2/MeV}$ corresponding to the largest
permissible value of
$\tau_c$.The peak is due to the fact that
the permissible values of $\tau_c$ are maximized when the neutrino is in
resonance somewhere between $0.7R_{\odot}$ and $0.85R_{\odot}$.
If one wished to satisfy the condition on $\tau_c$
for $\Delta m^2 / E$ less than or equal to $4\times 10^{-8} {\rm eV^2 /
MeV}$, one would obtain about 3 \% muon antineutrinos on average.
This is the same order of magnitude as that is produced by
resonant transitions in this region of the sun, for comparable magnetic
moments and magnetic fields \cite{sof,me,schram}.

We would also like to point out that in the case of non-zero mixing
between neutrino species in a purely random magnetic field
like that considered above, one can derive the exact (assuming Eq. (26) holds)
time dependence for the sum of the neutrino probabilities i.e. $|\psi_e(r)|^2
+ |\psi_{\mu}(r)|^2 $. The Hamiltonian is given by \cite{lam},
\begin{equation}
\label{eqn31}
     \hat H = \left(
        \begin{array}{cc}
             H_L & 0 \\
             0 & H_R
        \end{array}\right) + \mu B_r\left(
        \begin{array}{cc}
             0 & \hat M^{\prime} \\
             -\hat M^{\prime} & 0
        \end{array}\right),
\end{equation}
where, $\hat M^{\prime} = i\hat{\sigma}_y$ and,
\begin{equation}
\label{eqn32}
H_{L,R} = \left(
     \begin{array}{cc}
        -{{\Delta m^2}\over 4E}\cos 2\theta \pm a_e &
        {{\Delta m^2}\over 4E}\sin 2\theta \\
        {{\Delta m^2}\over 4E}\sin 2\theta &
        {{\Delta m^2}\over 4E}\cos 2\theta \pm a_{\mu}
     \end{array}\right).
\end{equation}
If one evaluates Eq. (16) one can derive the equation for the
sum of the neutrino probabilities. The solution is given by Eq. (23),
with $<|\psi_1(T)|^2>$ replaced by the sum of the electron and muon
neutrino probabilities. Therefore, the same estimates as in Fig. (3)
apply to this case but for the sum of the antineutrino averaged
probabilities.
One should note that the flavor precession
will, in general, be changed by the inclusion of the magnetic field.

\subsection{Flavor Precession with a Noisy Matter Density}

\indent

The equation governing matter-enhanced flavor oscillations (for
two flavors) is given by,
\begin{equation}
\label{eqn33}
i{d\over dr}\left(
        \begin{array}{c}
           \psi_{\nu_e} \\
           \psi_{{\nu}_{\mu}}
        \end{array}\right) = \left(
        \begin{array}{cc}
       -{{\Delta m^2} \over 4E}\cos 2\theta + {1\over \sqrt{2}}G_F N_e &
       {{\Delta m^2} \over 4E}\sin 2\theta \\
       {{\Delta m^2} \over 4E}\sin 2\theta &
       {{\Delta m^2} \over 4E}\cos 2\theta - {1\over \sqrt{2}}G_F N_e
        \end{array}\right)\left(
        \begin{array}{c}
            \psi_{\nu_e} \\
            \psi_{\nu_{\mu}}
        \end{array}\right),
\end{equation}
where $\theta$ is the vacuum mixing angle. We investigate the case in which
$Ne = (1 + \beta)\overline{N_e}(r)$ where $\beta$ is a random quantity whose
rms value is
the rms value of the density fluctuations relative to the average
density and $\overline{N_e}$ is the electron density in the standard
solar model. We satisfy Eq. (26) by choosing,
\begin{equation}
\label{eqn34}
\tau_c = 0.1 \times ({{\Delta m^2} \over 2E}\sin 2\theta)^{-1},
\end{equation}
where the logarithmic derivative in Eq. (26) is much smaller for
the sun than the term we have kept.
We numerically solve Eq. (24) for the electron neutrino survival
probability (averaged for infinite distance beyond the surface of the sun),
as a function of
$\Delta m^2 / E$ for fixed values of $\sin^2 2\theta$ and
$\sqrt{<\beta^2>}$, with $\tau_c$ given by Eq. (35). We have used the solar
electron density of Bahcall and collaborators. Figures (4a) through (4d) show
the electron neutrino survival probability as a function of
$\Delta m^2 / E$ for $\sin^2 2\theta = 0.8, 0.1, 0.01$ and $0.001$
respectively, with $\sqrt{<\beta^2>} = 0.02$ In these figures, the curve
which reaches the smallest values in the center of the plot (from about
$\Delta m^2 / E = 1 \times 10^{-6}$ to $1 \times 10^{-5} {\rm eV^2 /
MeV}$) is the probability in the absence of the density fluctuations. One
observes a suppression of the flavor transition in this region by as much as
20 \% in Fig. (4c). The region of
the largest effect (around $\Delta m^2 / E = 1 \times 10^{-5}$ is
in agreement with the maximum value of $2k\Delta r$, where $\Delta r$ is
the width of the resonance region. In order to show the sensitivity of the
suppression to the size of the fluctuations, we show the same plots as Fig.
(4a)-(4d) in Fig. (5a)-(5d), with the exception that the rms fluctuation
in the density is doubled to 4 \% of the average density. One observes
a very large effect in the previously mentioned region, which saturates
at a probability of one half, for values near $\Delta m^2 / E = 1
\times 10^{-5} {\rm eV^2/MeV}$.
One should note that in Figures (4a)-(5d), $\tau_c$ was chosen for each
value of $\Delta m^2 / E$ to obey Eq. (35). Thus, the $\tau_c$
chosen for $\Delta m^2 / E = 1 \times 10^{-5} {\rm eV^2 / MeV}$,
while permissible
for smaller values of $\Delta m^2 / E$, would not have given as
large an effect, and would, for larger values of $\Delta m^2 / E$
violate Eq. (35).

In Figs. (4) and (5), there appears to be an enhancement of the transition
probability for values of $\Delta m^2/E > 1.5 \times 10^{-5}$. Neutrinos of
with these parameter values do not go through a resonant transition in
the sun, since a resonant transition would require a density larger than
the maximum solar density. However, to a varying extent they are within the
second half of their resonant region near the center of the sun. The solar
density profile near the center is rather flat and therefore one obtains
an enhancement of the transition probability similar to that in Fig. (3c)
and (3d).

One needs to question whether such fluctuations can arise
in the sun, and whether the correlation lengths, which can be quite small
($\tau_c$ for $\Delta m^2 / E = 1 \times 10^{-5} {\rm eV^2 / MeV}$ and
$\sin^2 2\theta
= 0.01$ is about $10$km ) could be realistic.
If one were to do a numerical study
utilizing a good random number generator, one would not be limited by
Eq. (35) and could consider any correlation length. Our purpose is only
to show that such effects {\it could be} important.

\section{Conclusions}

\indent

We have derived a Redfield
differential equation for the time dependence of averaged
values of functions of the probability of finding an N level system in the
N$^{th}$ level after being subjected to a randomly fluctuating field, of the
delta-correlated gaussian type. This formalism applies as an approximation
to the case where the fluctuating field can be described by  a finite
correlation time. This approximation is valid if the product of the correlation
time and the energy scale of the Hamiltonian is small. Upon applying this
to neutrino flavor or spin-flavor precession, we have shown that the
probability will relax eventually to a value of one half, if the neutrino
spends a sufficiently long time in a medium with randomly fluctuating
matter density or a randomly fluctuating magnetic field. In the case of
a fluctuating magnetic field, the relaxation is independent of the
matter density, assuming the correlation time of the field fluctuations
is small compared to the neutrino oscillation length in matter.
In the case of fluctuations added to a constant matter density, the probability
again will eventually relax to a value of one half, but the relaxation time
is greatly increased if one is far from the resonant condition.

We have also examined the case of neutrino spin-flavor precession
in the sun, for a purely random magnetic field and no flavor mixing. It
appears that the current limits on the neutrino magnetic moment and
the guesses concerning the maximum values of the solar magnetic field, combine
to give only a small effect on the average electron neutrino flux.
When there is flavor mixing and a
purely random field, the combined
average probabilities of the neutrinos is again a simple exponentially
decreasing function and therefore the results of a purely random field
and no flavor mixing apply to this case as well.
For the case of a randomly fluctuating electron
density, the MSW effect can be strongly suppressed for rms fluctuations
of 4 \% of the local electron density. However, this requires correlation
lengths of about 40km, and seems to give a significant effect only for
neutrinos which have their MSW resonant transition deep in the sun.

In spite of these problems, we believe that a numerical study, which should
give similar results for correlation times which do not badly violate Eq. (35),
may bring to light many interesting effects. Implications of the
density fluctuations discussed here on stellar collapse and supernova
dynamics will be published elsewhere \cite{fl}.
\vskip .2in
\centerline{\bf ACKNOWLEDGMENTS}
\vskip .2in
We thank to G. Fuller, W. Haxton, and Y. Qian for very useful discussions.
This research was supported in part by the U.S. National Science
Foundation Grant No. PHY-9314131 and in part by the University of
Wisconsin Research Committee with funds granted by the Wisconsin Alumni
Research Foundation. F. N. L.'s research was supported
in part by a grant from Mr. E. J. Loreti.
We would also like to thank the Institute for Nuclear Theory at the
University of Washington for its hospitality and the Department of
Energy for partial support during the completion of this work.
\vskip .2in
\centerline{\bf Appendix }
\vskip .2in
\indent

For $\hat H_0(t)$ given in Eq. (5), one selects the element of largest value
during the time the system is in the presence of the random field. Let
this largest value be $E_{max}$. We consider the case where $ \tau E_{max}
\ll 1$. We take $f_{ij}$ in Eqs. (10) and (11) to be given by,
\begin{equation}
f_{ij} = \theta( \tau - |t_i - t_j| ),
\end{equation}
where again the average of odd products vanish. We rewrite the average of
Eq. (9) as,
\begin{equation}
<\hat {\rho}_I(t)> = \hat {\rho}_0 + <\hat {\rho}_I^{(2)}(t)> + <\hat
{\rho}_I^{(4)}(t)> +
<\hat {\rho}_I^{(6)}(t)> + \cdots
\end{equation}
where,
\begin{eqnarray}
<\hat {\rho}_I^{(2N)}(t)> &=&  (-1)^N \alpha^{2N} \int_0^t dt_{2N}
\int_0^{t_{2N}} dt_{2N-1}\cdots \nonumber \\
&\phantom{=}&+ \int_0^{t_1} dt_1 F_{2N} [\hat M(t_{2N}),
[\hat M(t_{2N-1}), \cdots \hat M(t_1)]\cdots ]]
\end{eqnarray}
and,
\begin{equation}
F_{2N} = \sum_{n_1\cdots n_{2N}}^{P(1,2,\cdots ,2N)} f_{n_1n_2}f_{n_3n_4}
f_{n_5n_6}\cdots f_{n_{2N-1}n_{2N}},
\end{equation}
where, P(1,2,$\cdots$ ,2N) means all
permutations. We explicitly show the second and third terms:
\begin{eqnarray}
&\phantom{=}&<\hat {\rho}_I^{(2)}(t)>\ \sim -\alpha^2 \int_0^t dt_1 [\hat
M(t_1),
\int_{t_1-\tau}^{t_1}[\hat M(t_1) + {d\hat M(t_1) \over dt}
(t_2 - t_1), \hat {\rho}_0]] \phantom{oiueoe} \nonumber \\
&\sim & -\alpha^2 \tau \int_0^t dt_1 [\hat M(t_1),[\hat M(T_1), \hat {\rho}_0]]
+ \alpha^2 {\tau^2 \over 2} \int_0^t dt_1 [\hat M(t_1), [{d\hat M(t_1) \over
dt}
, \hat {\rho}_0]],
\end{eqnarray}
where $ d\hat M(t_1)/dt = i \hat U_0^{\dagger} [\hat H_0, \hat
M^{\prime}] \hat U_0 $
and is
therefore proportional to $E_{max}$. In the third term one has a sum of
three products of two f's, ($ f_{12}f_{34} + f_{13}f_{24} +
f_{14}f_{32}$) only the first of which has the times in the order of the
times appearing in the nested integrals. The first of these three terms gives,
\begin{eqnarray}
<\hat {\rho}_I^{(4)}(t)>_{12,34} &\sim& \alpha^4 \tau^2
\int_0^t dt_1 [\hat M(t_1), \bigl\{ [\hat M(t_1), \int_0^{t_1} dt_2 [\hat
M(t_2),
[\hat M(t_2),\hat {\rho}_0]]]] \phantom{oioo} \nonumber \\
&-& {\tau \over 2}\bigl( [{d\hat M(t_1) \over dt},\int_0^{t_1}dt_2
[\hat M(t_2), [\hat M(t_2), \hat {\rho}_0]]]] \nonumber \\
&+& [\hat M(t_1),\int_0^{t_1}dt_2
[\hat M(t_2),[{d\hat M(t_1) \over dt}, \hat {\rho}_0]]]] \\
&\phantom{+}& + [\hat M(t_1),[\hat M(t_1),[\hat M(t_1),\hat
{\rho}_0]]]] \bigr) \bigr\} +
O(\tau^4). \nonumber
\end{eqnarray}
The last term, however, is not proportional to $E_{max}$. This will
cause a problem since it could contribute to the next order term of the average
. It turns out that it cancels the largest term coming from the remaining two
f products. That these remaining f-products are of largest order $\tau^3$ can
be seen by noting that when the argument of the theta function in an f connects
two non-sequential times, the intermediate time(s) must also be within $\tau$
of the larger time in the theta function. For example, the remaining two
contributions to $<\hat {\rho}_I^{(4)}(t)>$ are,
\begin{eqnarray}
<\hat {\rho}_I^{(4)}(t)>_{13,24} &+& <\hat {\rho}_I^{(4)}(t)>_{14,23} =
\phantom{oiueoi}\nonumber \\
&\phantom{=}& \alpha^4 \int_0^t dt_1 [\hat M(t_1),[\hat M(t_1),[\hat
M(t_1),[\hat M(t_1),
\hat {\rho}_0]]]] \nonumber \\
&\times&
\int_{t_1-\tau}^{t_1}
dt_2 \int_{t_1-\tau}^{t_2} dt_3 \Bigl( \int_{t_2-\tau}^{t_3} dt_4  +
\int_{t_1-\tau}^{t_3} dt_4 \Bigr) \\
&=& \alpha^4 {\tau^3 \over 2} \int_0^t
dt_1 [\hat M(t_1),[\hat M(t_1),[\hat M(t_1),[\hat M(t_1),\hat
{\rho}_0]]]]. \nonumber
\end{eqnarray}
This feature
appears to continue throughout each term in the entire expression.
Therefore, if $\tau E_{max} \ll
1$, and neglecting terms of order $\tau E_{max}$ and smaller,
\begin{eqnarray}
<\hat {\rho}_I^{(2N)}(t)> &=& (-1)^N \alpha^{2N} \tau^N \int_0^t dt_{2N}
[\hat M(t_{2N}),[\hat M(t_{2N}), \phantom{oiueoiueoieoie}\nonumber \\
&\phantom{=}& \int_0^{t_{2N}} dt_{2N-2}
[\hat M(t_{2N-2},[\hat M(t_{2N-2}), \int_0^{t_{2N-2}}
\cdots \\
&\phantom{+}& \int_0^{t_4} dt_2 [\hat M(t_2),[\hat M(t_2), \hat
{\rho}_0]]\cdots ]]]].
\nonumber
\end{eqnarray}
which leads to Eqs. (15) \& (16).

In the above derivation we have assumed $\alpha $ to be a constant. If
$\alpha$ depends on time, the condition, $\tau E_{max} \ll 1$ , should be
replaced by the condition,
\begin{equation}
\tau [ (d/dt \log \alpha(t))^2 + E_{max}^2]^{1\over 2} \ll 1.
\end{equation}
\vfill
\eject

{}
\vfill \eject

\vskip .2in
\centerline{\bf Figure Captions}
\vskip .2in
\noindent
{\bf Figure 1} $<|\psi_1(t)|^2>$ obtained by numerical solution of Eq.
(21) (randomly fluctuating off-diagonal) with $B_0 = 1$, $k = 0.05$ for
(a) $A = 0$, (b) $A = 0.5$, (c) $A=2.0$ and (d) $A = 10$. The initial
condition is $\psi_1(0) = 1, \psi_2(0) = 0$.
\vskip .2in
\noindent
{\bf Figure 2} $<|\psi_1(t)|^2>$ obtained by numerical solution of Eq.
(24) (randomly fluctuating diagonal) with $B_0 = 1$, $k = 0.05$ for
(a) $A = 0$, (b) $A = 0.5$, (c) $A=2.0$ and (d) $A = 10$. The initial
condition is $\psi_1(0) = 1, \psi_2(0) = 0$.
\vskip .2in
\noindent
{\bf Figure 3} $k \Delta r$ as a function of
$\Delta m^2 / E$ for a randomly fluctuating magnetic field of rms
value $1 \times 10^5 G$ confined to a region of the solar convective
zone, as described in the text.
\vskip .2in
\noindent
{\bf Figure 4} $<|\psi_e|^2>$ for infinite distance as a function of
$\Delta m^2 / E$. The lower(upper) curve around $\Delta m^2 / E = 1 \times
10^{-5}$ is the curve for a noiseless (noisy) electron density. The
rms value of the randomly fluctuating noise is 2 \% of the local
density. $\sin^2
2\theta$ is changed from (a) 0.8, (b) 0.1, (c) 0.01 and (d) 0.001.
 The
correlation length is given by Eq. (35) for each value of $\Delta m^2 /
E$.
\vskip .2in
\noindent
{\bf Figure 5} The same as Fig. (4) except that the rms value of the
randomly fluctuating density is 4 \% of the local density.
\vfill
\eject

\epsfysize=9in \epsfbox[95 25 550 675]{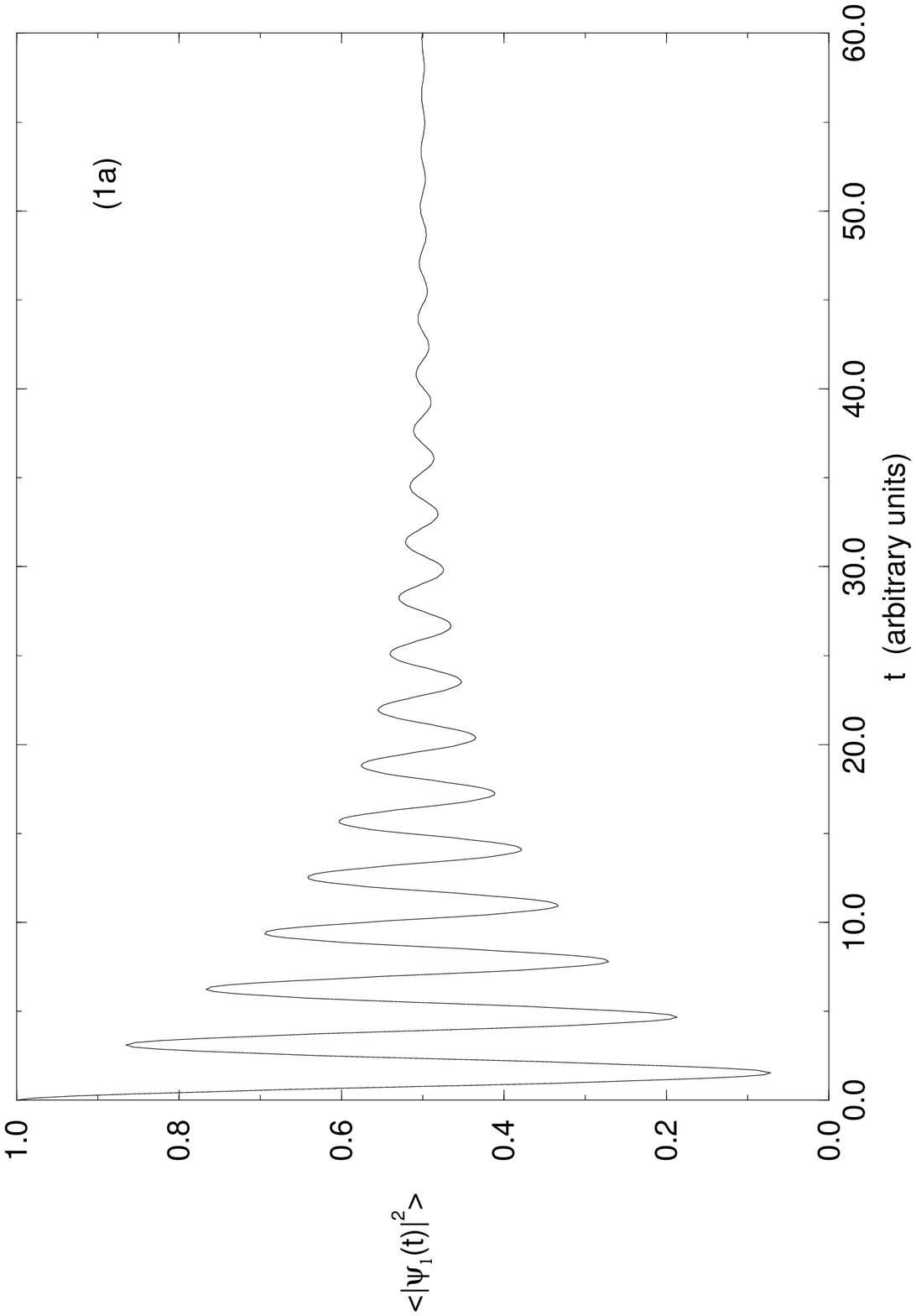}
\epsfysize=9in \epsfbox[95 25 550 675]{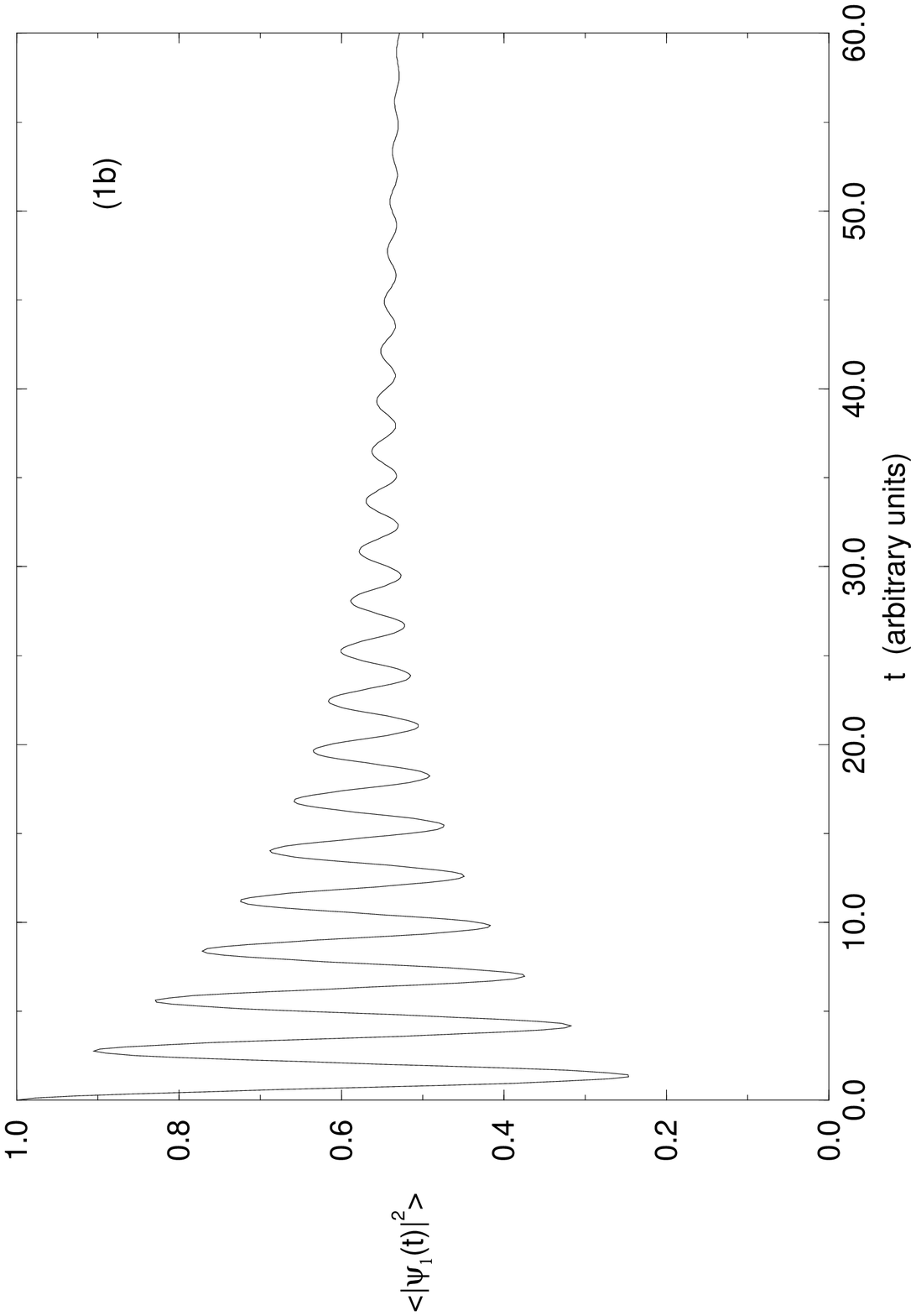}
\epsfysize=9in \epsfbox[95 25 550 675]{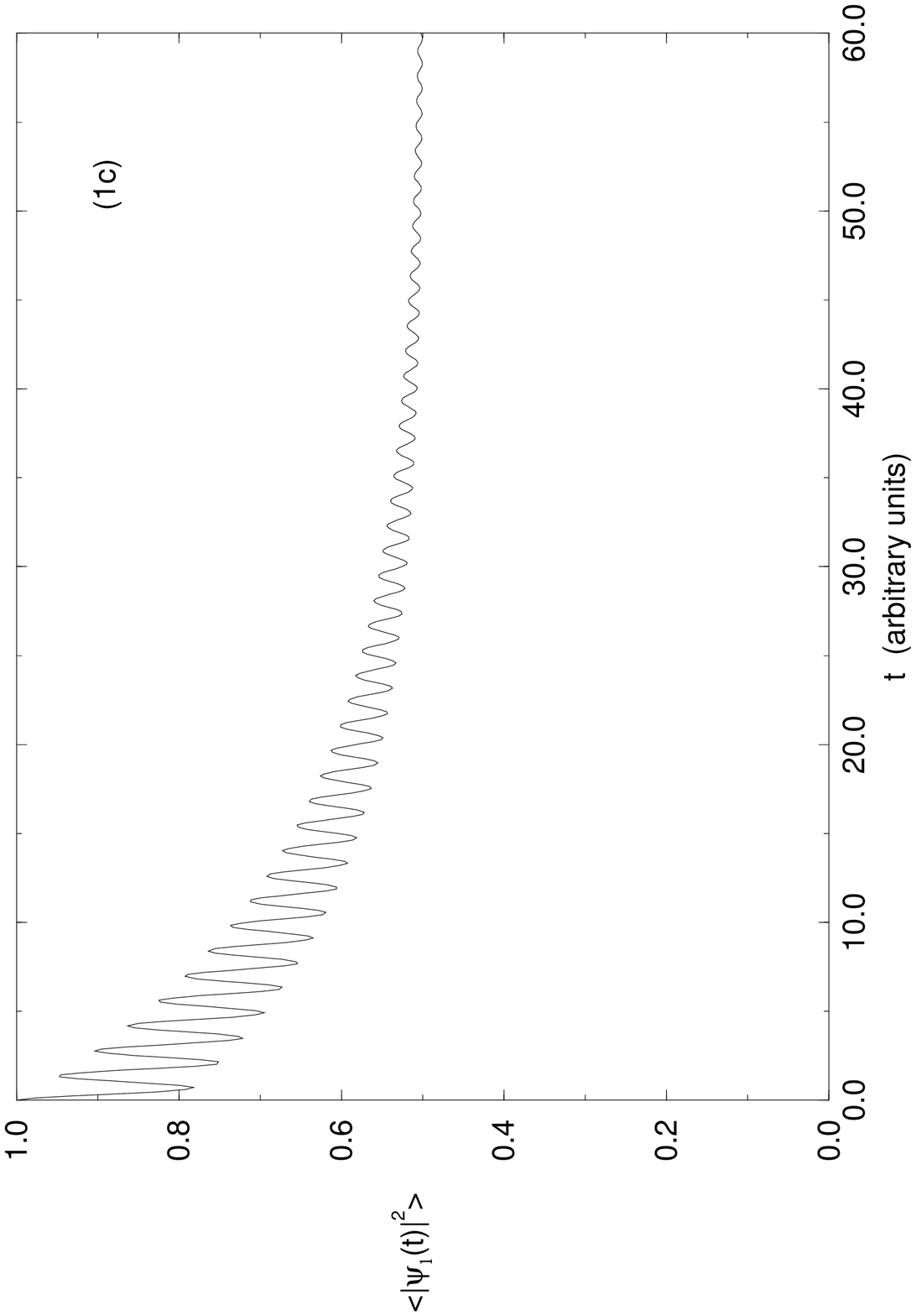}
\epsfysize=9in \epsfbox[95 25 550 675]{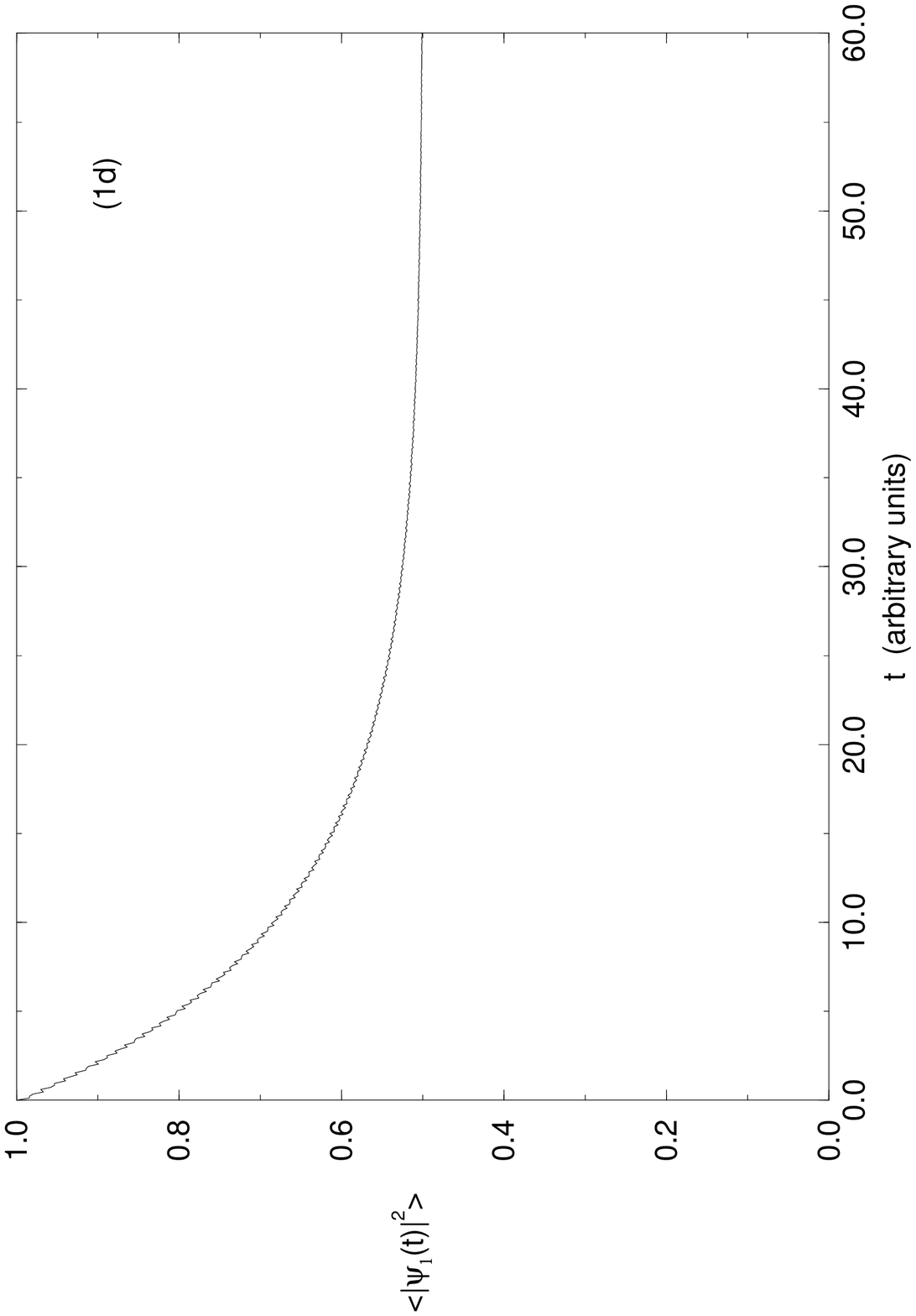}

\epsfysize=9in \epsfbox[95 25 550 675]{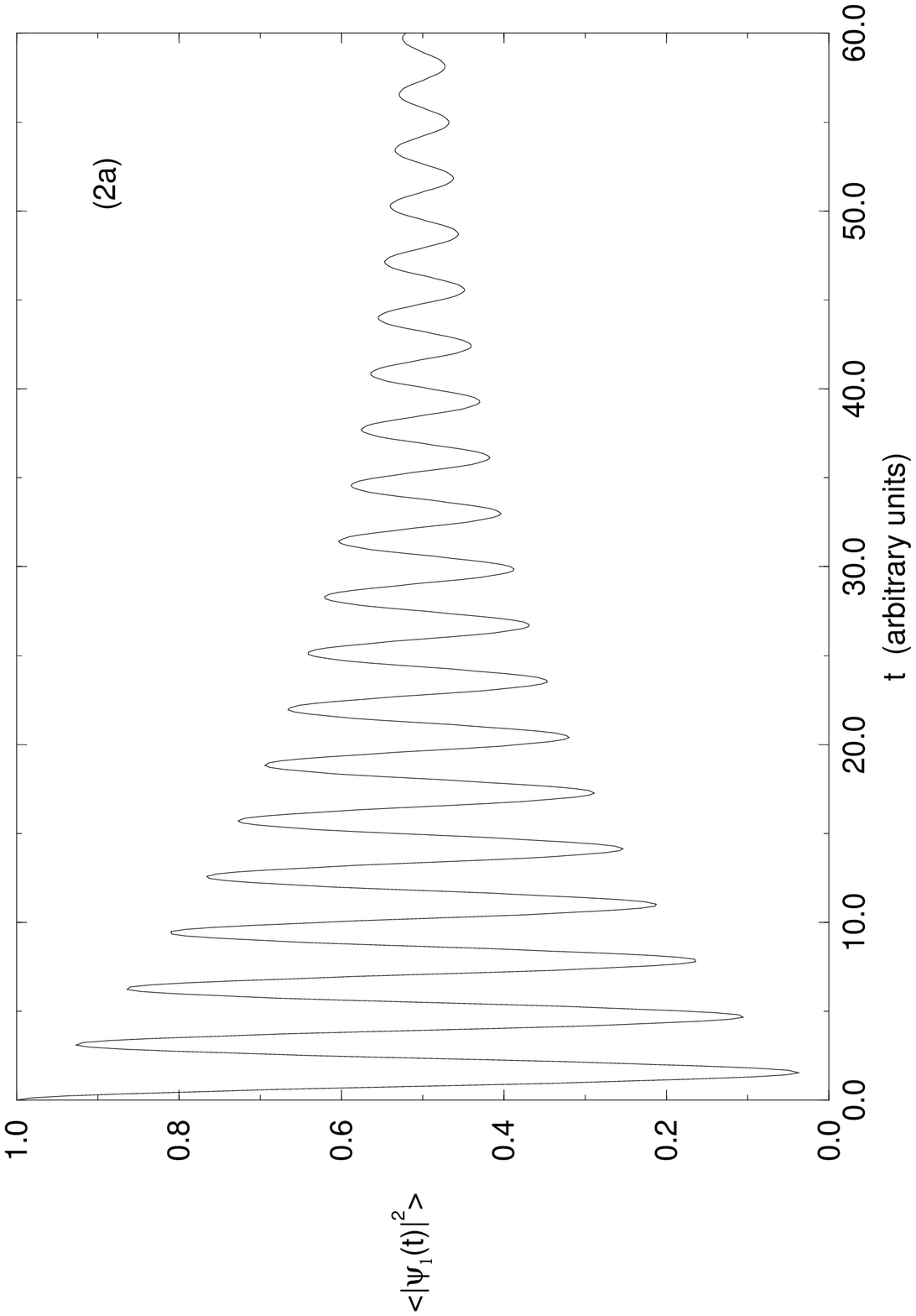}
\epsfysize=9in \epsfbox[95 25 550 675]{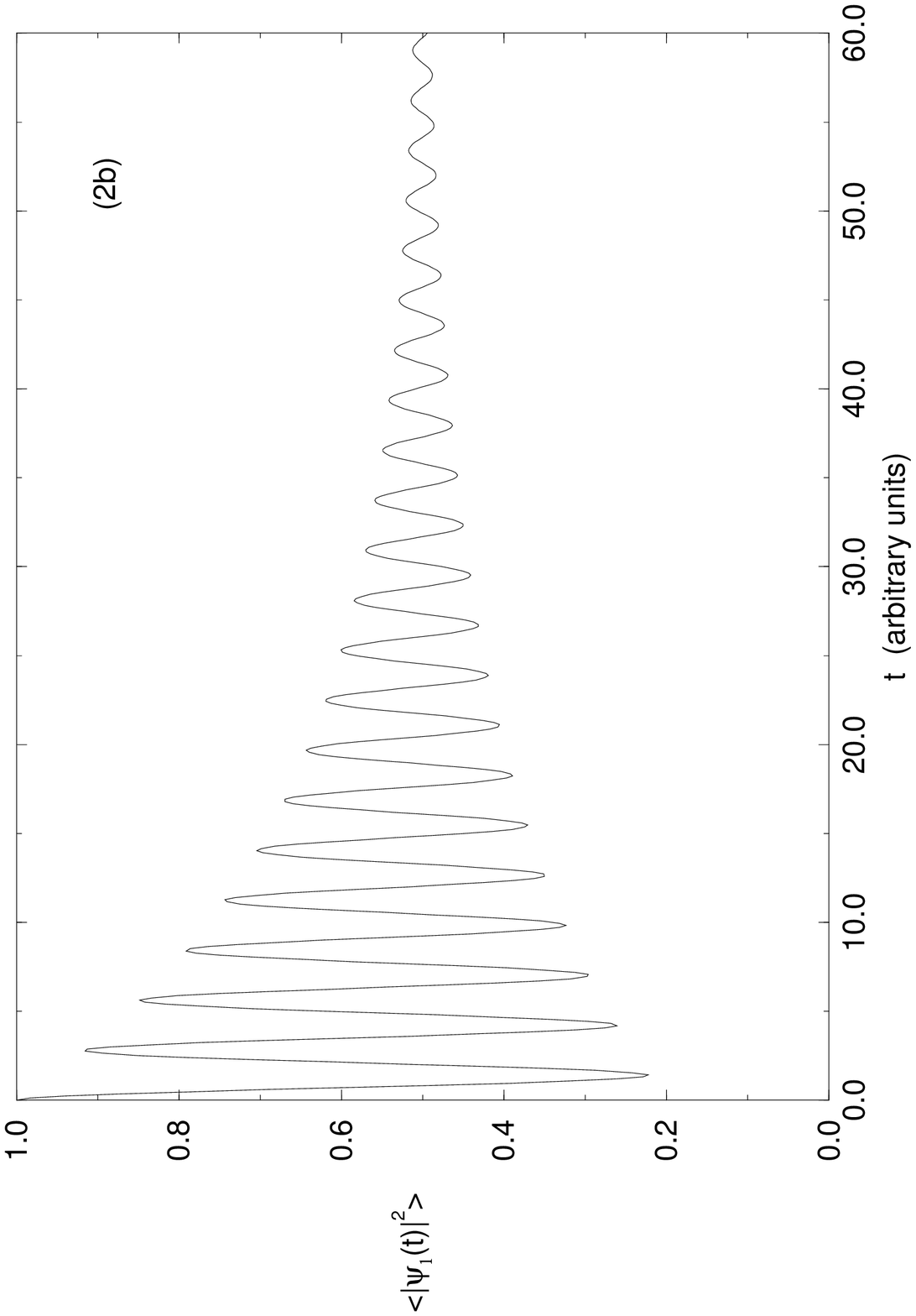}
\epsfysize=9in \epsfbox[95 25 550 675]{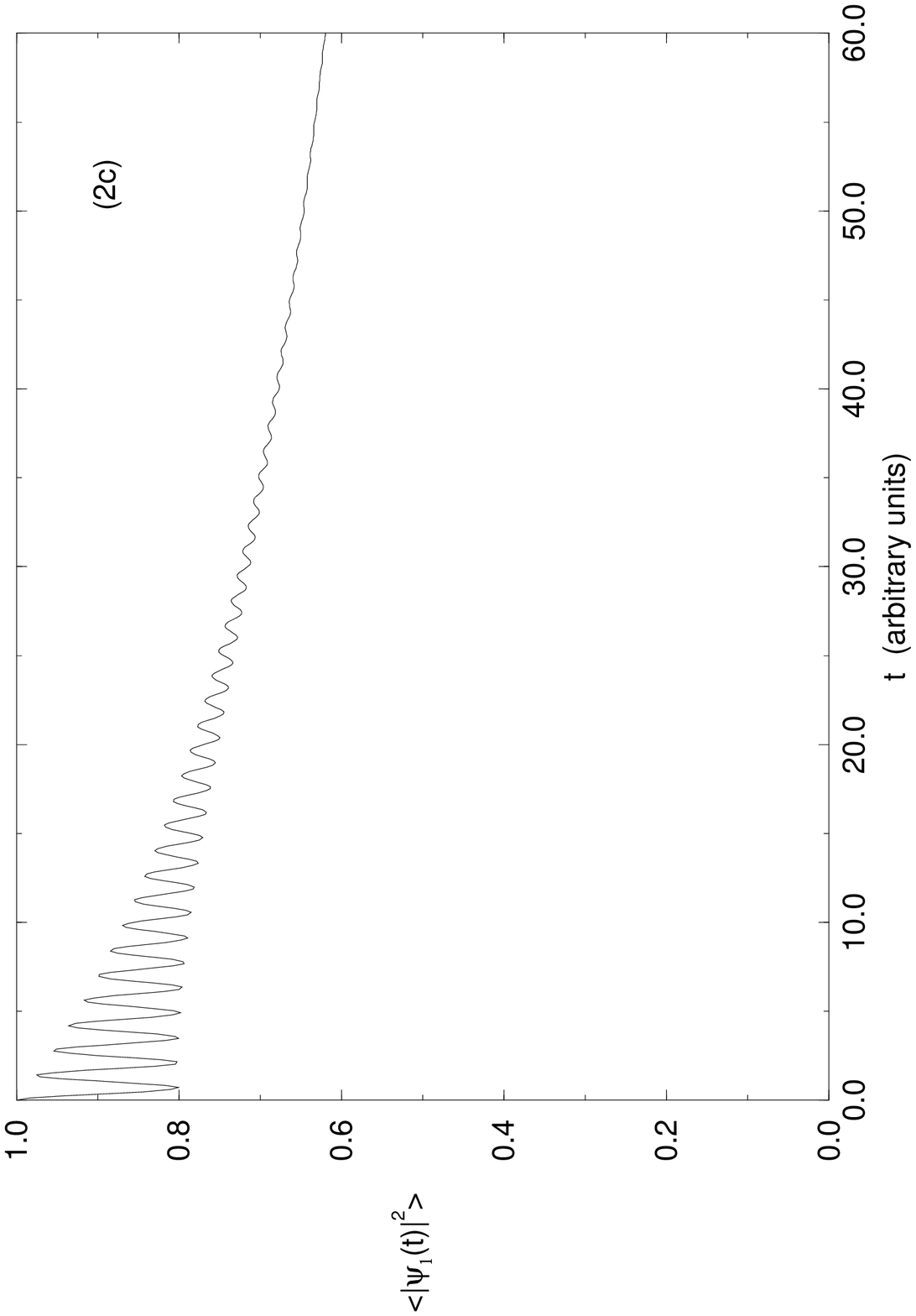}
\epsfysize=9in \epsfbox[95 25 550 675]{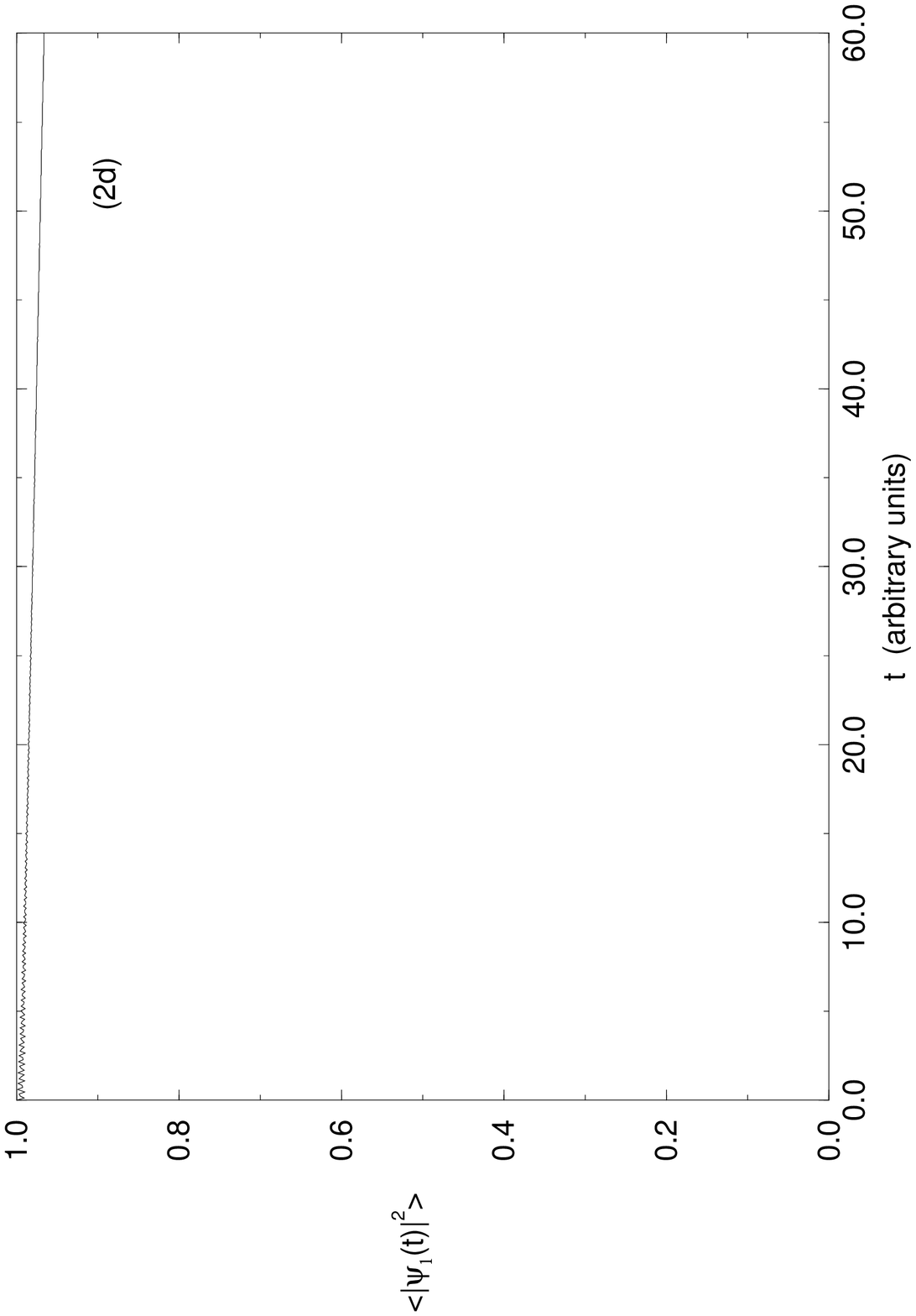}

\epsfysize=9in \epsfbox[95 25 550 675]{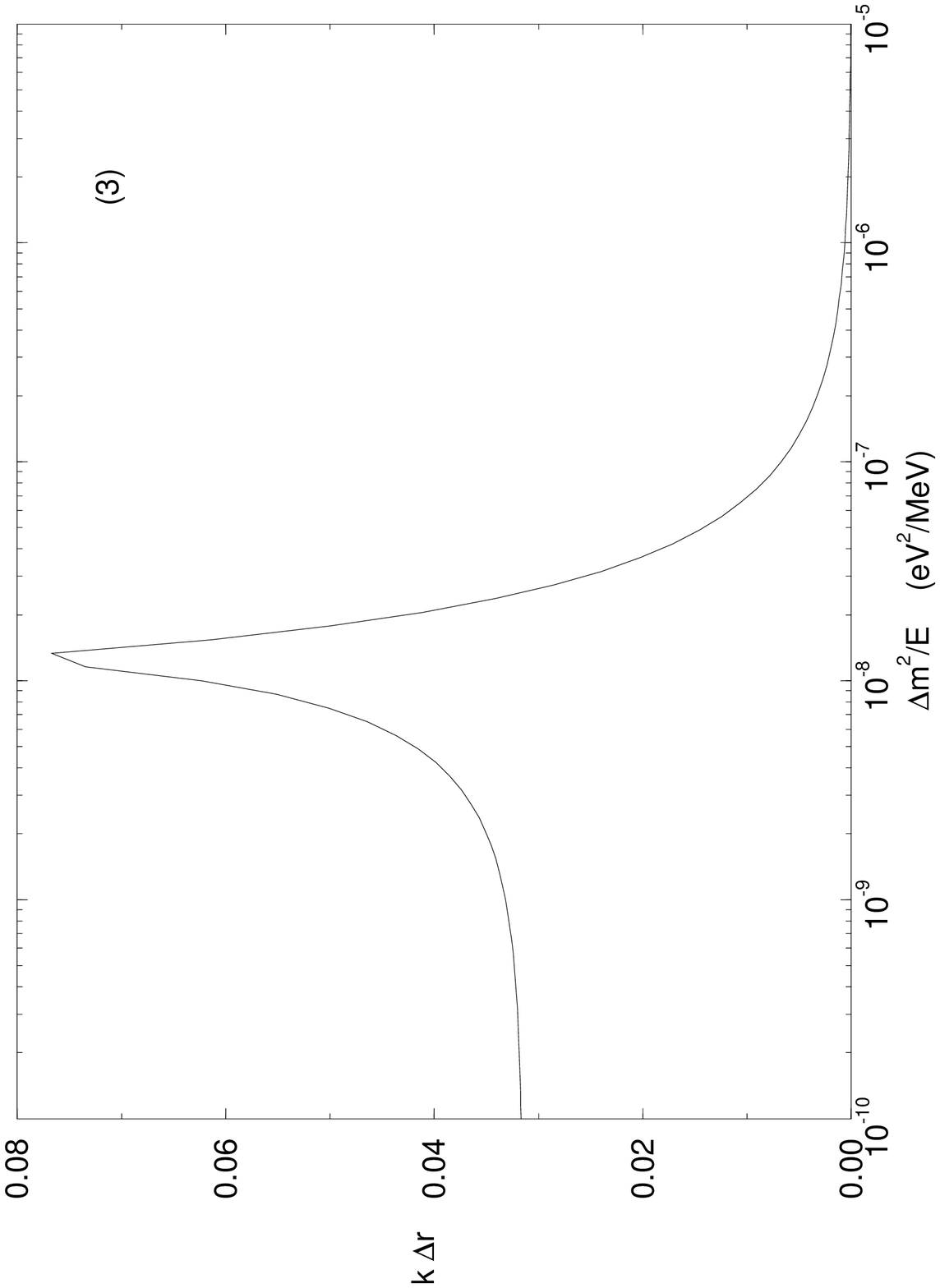}

\epsfysize=9in \epsfbox[95 25 550 675]{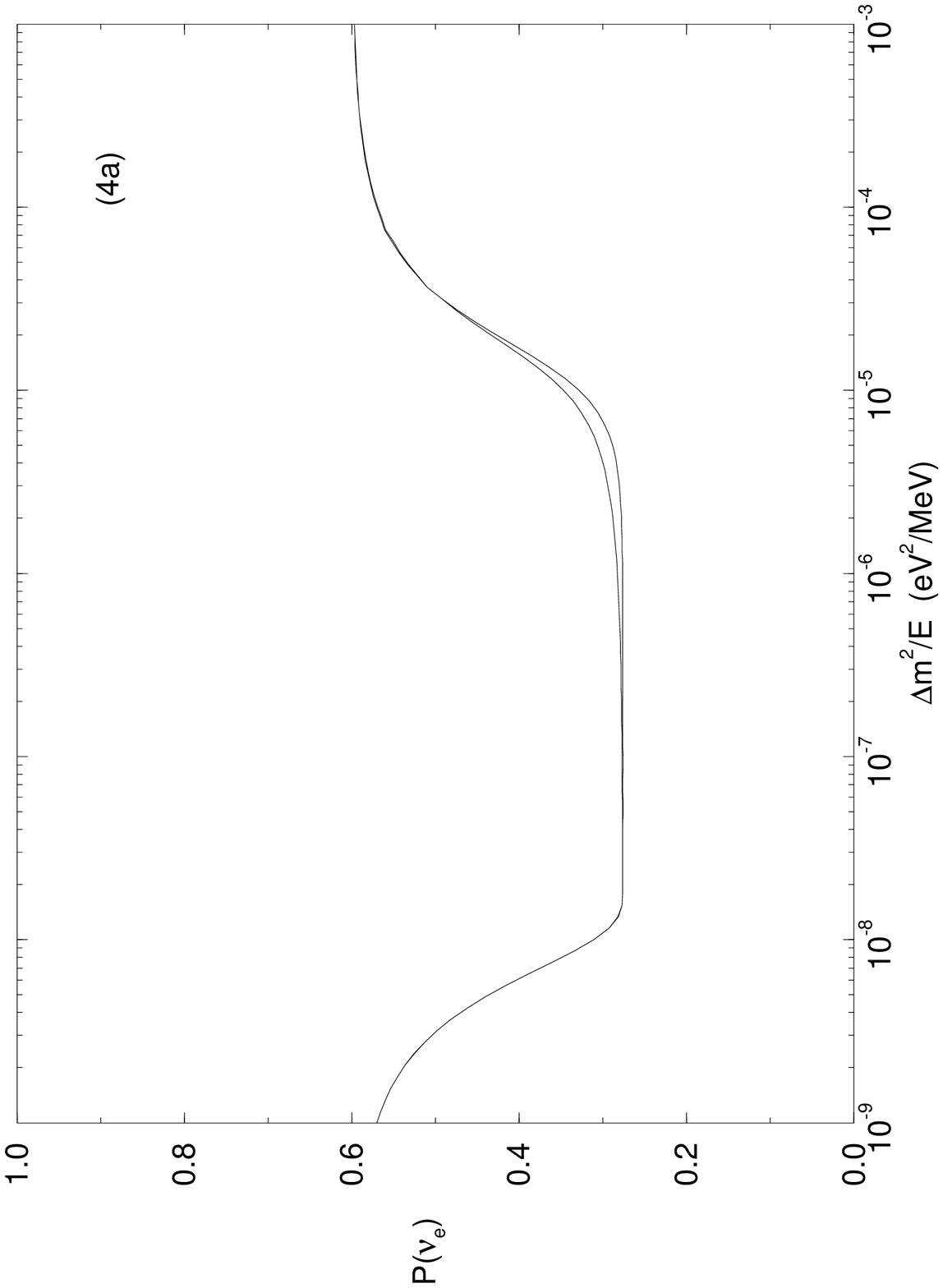}
\epsfysize=9in \epsfbox[95 25 550 675]{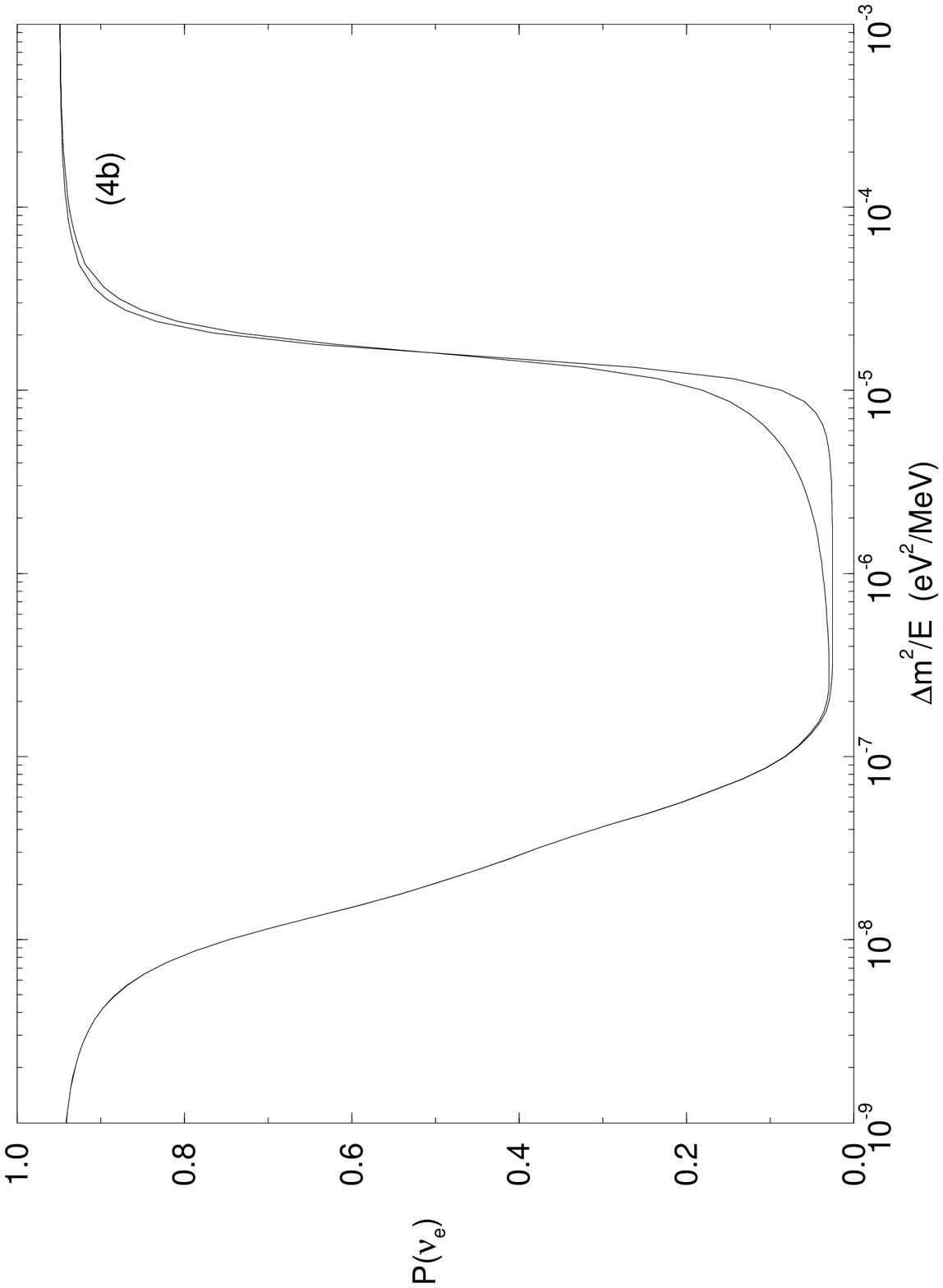}
\epsfysize=9in \epsfbox[95 25 550 675]{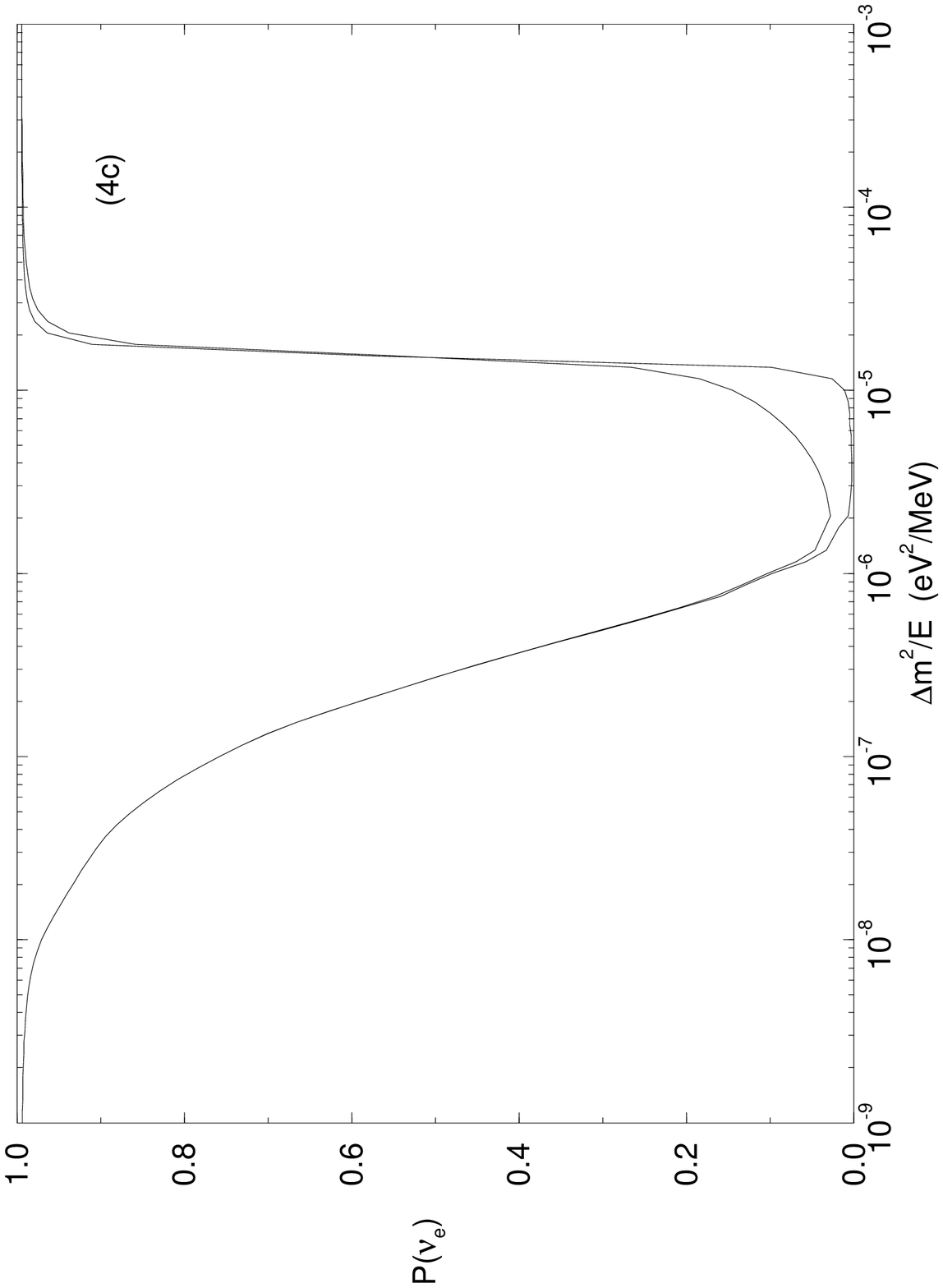}
\epsfysize=9in \epsfbox[95 25 550 675]{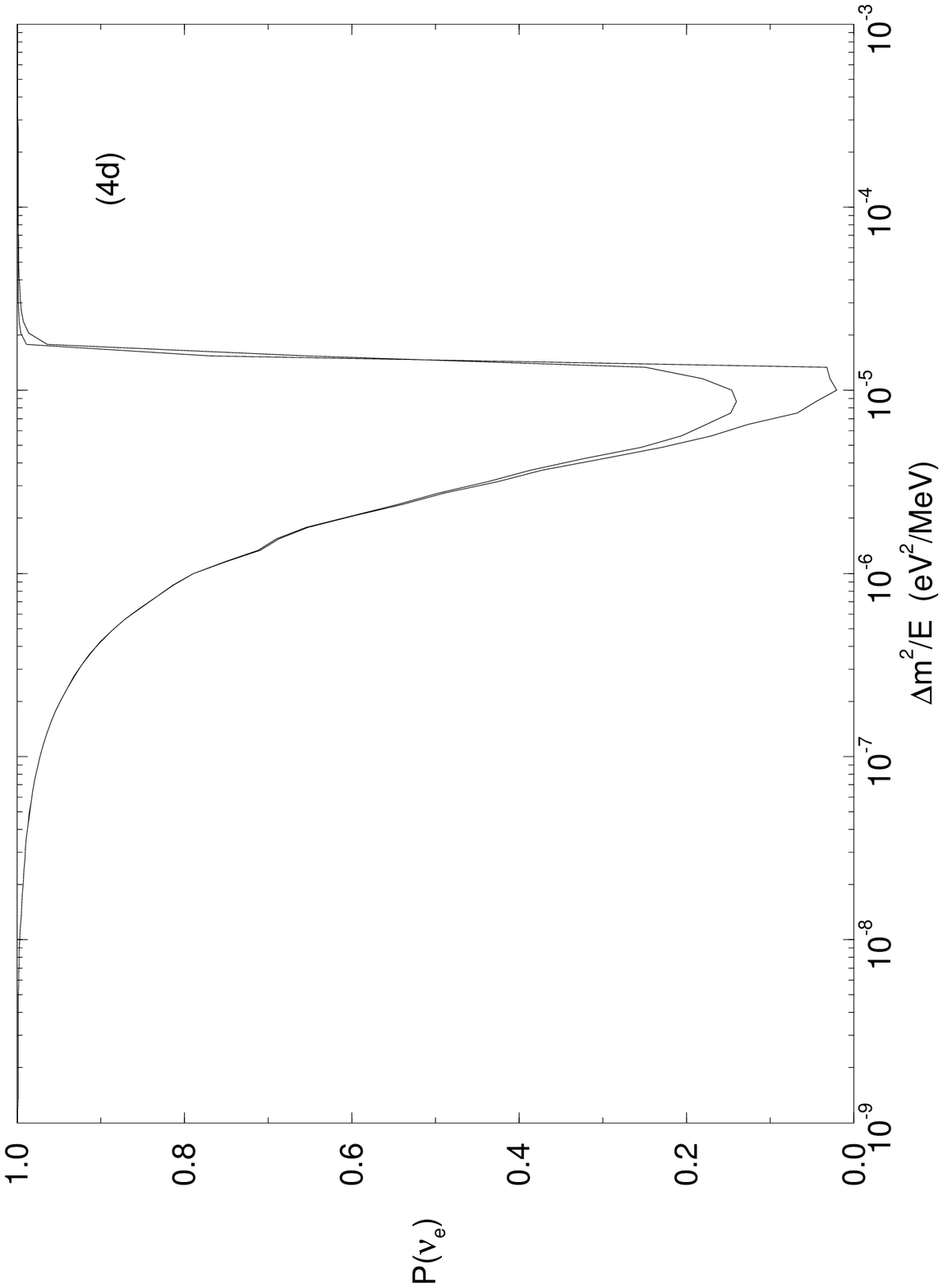}

\epsfysize=9in \epsfbox[95 25 550 675]{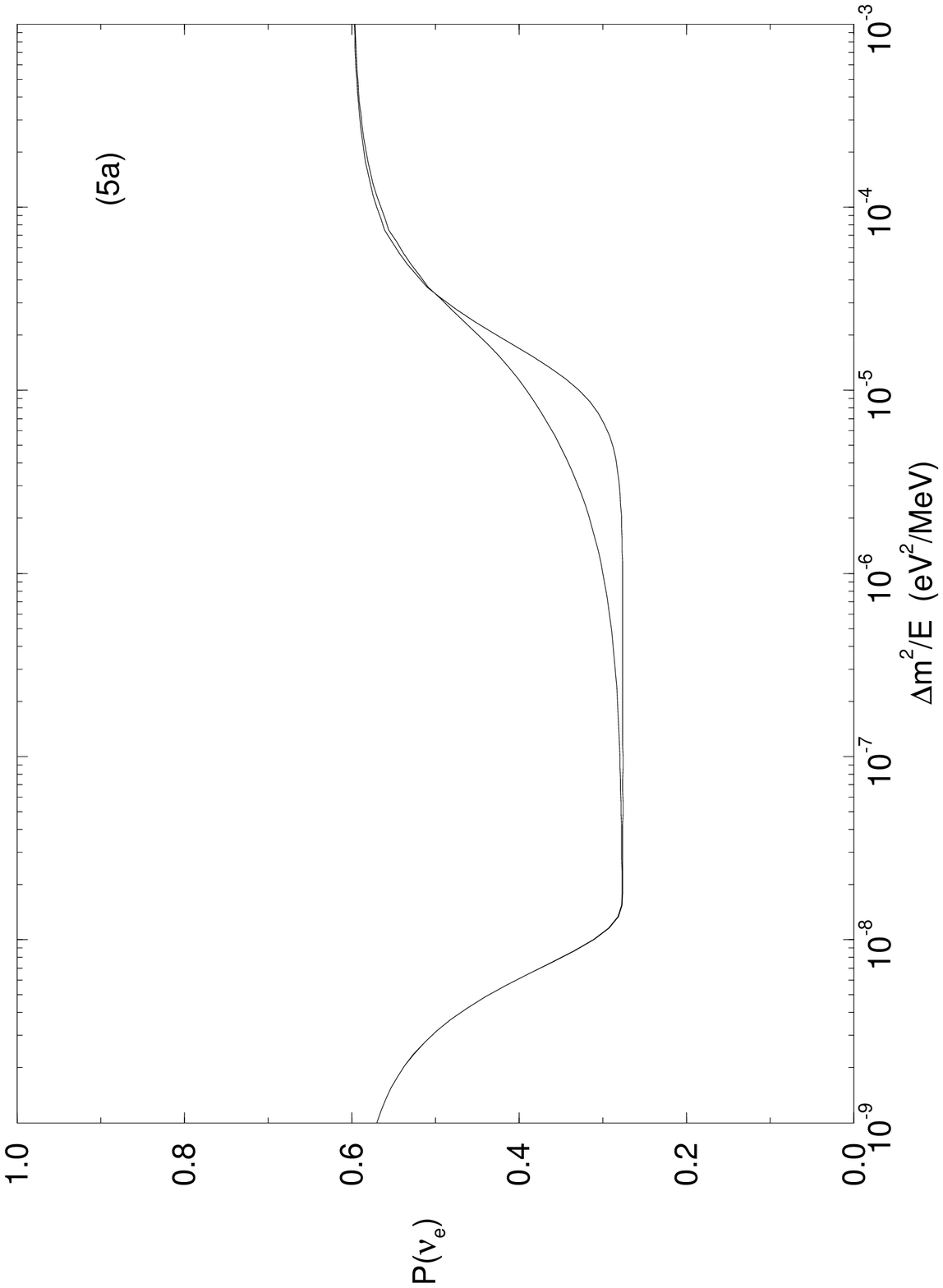}
\epsfysize=9in \epsfbox[95 25 550 675]{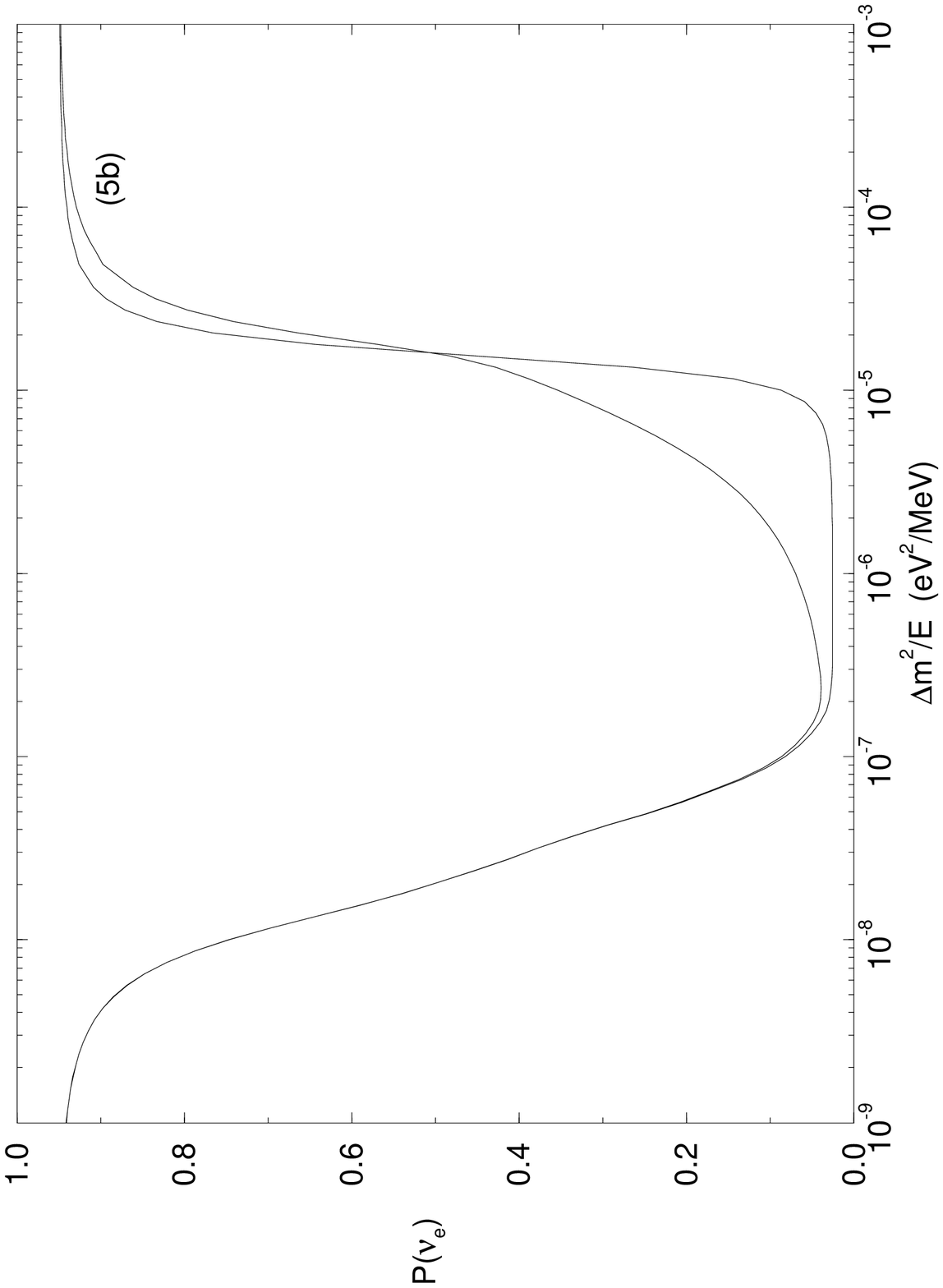}
\epsfysize=9in \epsfbox[95 25 550 675]{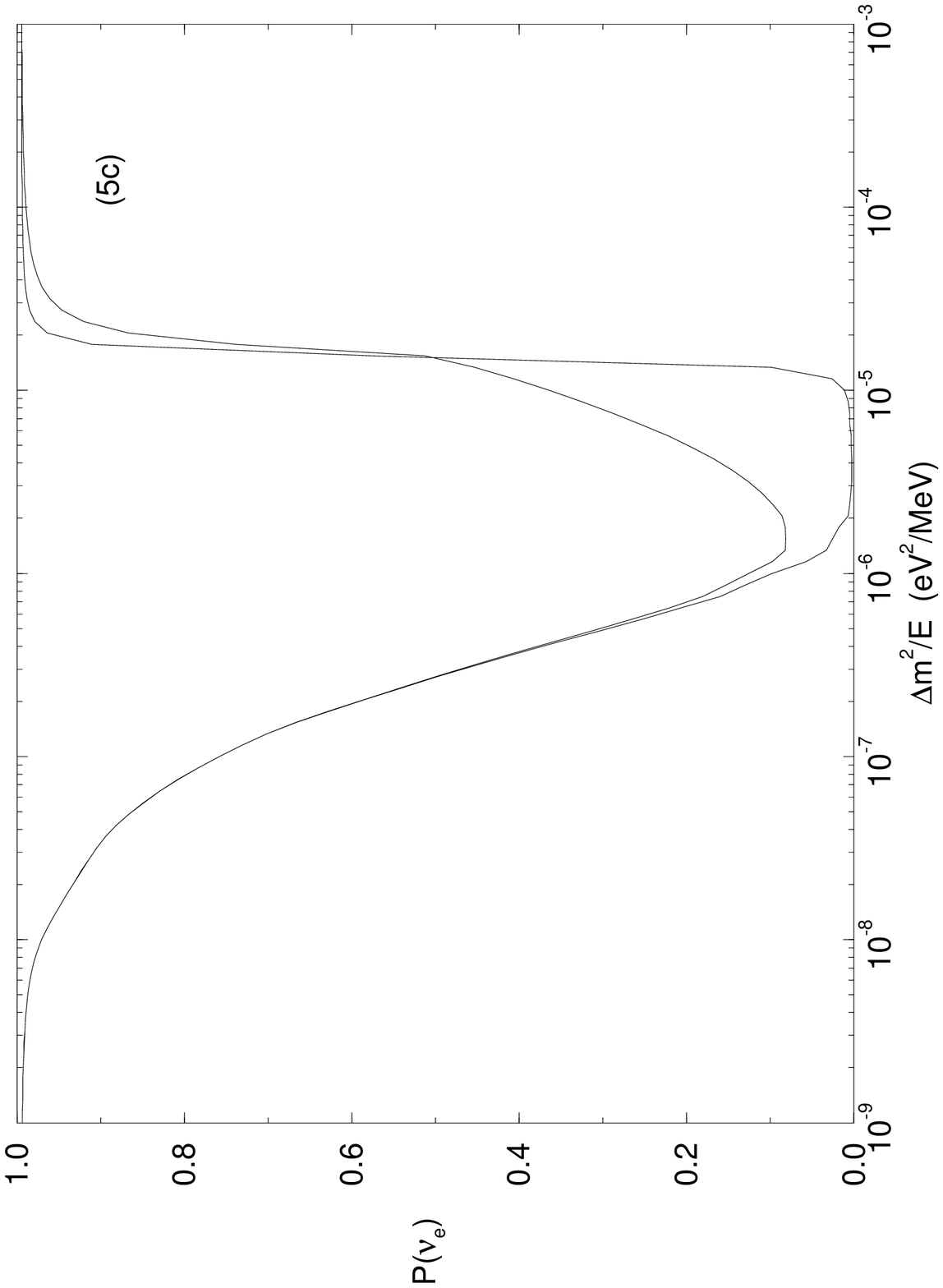}
\epsfysize=9in \epsfbox[95 25 550 675]{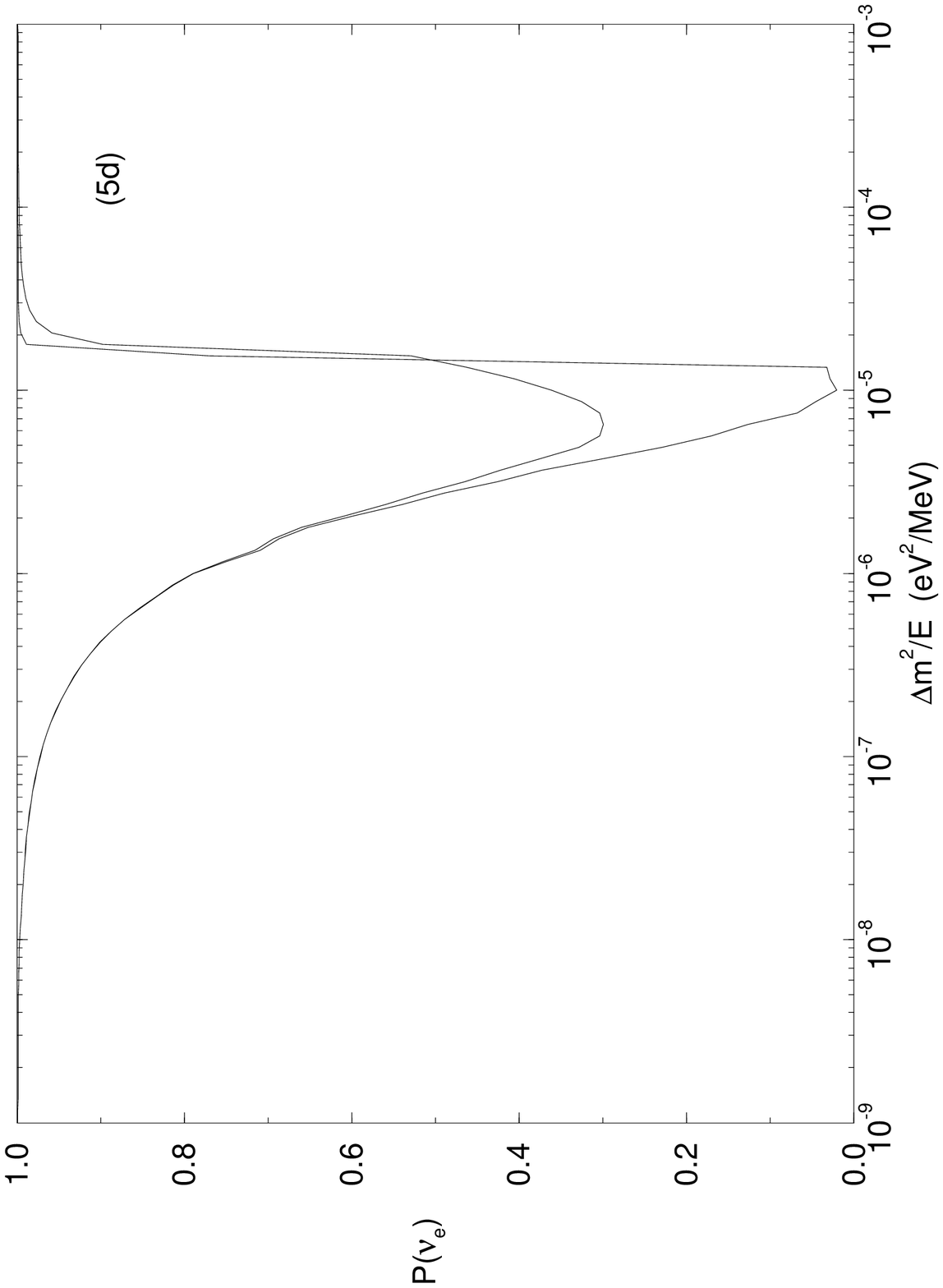}


\begin{thebibliography}{99}

\bibitem{msw}
L. Wolfenstein, Phys.\ Rev.\ {\bf D17}, 2369 (1978); {\bf D20}, 2634
(1979). S.P. Mikheyev and A. Yu. Smirnov, Nuovo Cim. {\bf 9C}, 17
(1986); Sov.  J. Nucl. Phys.\ {\bf 42}, 913 (1986).
\bibitem{us}
W.C. Haxton, Phys. Rev. Lett. {\bf 57}, 1271 (1986); S.J. Parke, {\it
ibid.} {\bf 57}, 1275 (1986); S.P. Rosen and J.M. Gelb, Phys. Rev. D
{\bf 34}, 969 (1986); A.B. Balantekin, S.H. Fricke, and P.J. Hatchell,
Phys. Rev. D {\bf 38}, 935 (1988); J.N. Bahcall and H.A. Bethe, Phys.
Rev. Lett.{\bf 65}, 2233 (1990).
\bibitem{lam}
C.-S. Lim and W.J. Marciano, Phys.\ Rev.\ {\bf D37}, 1368 (1988).
E.Kh. Akhmedov, Phys.\ Lett.\ {\bf B213}, 64 (1988); E.Kh. Akhmedov and M.
Yu. Khlopov, Mod.\ Phys.\ Lett.\ {\bf A3}, 451 (1988).
\bibitem{pant}
G. M. Fuller, R.W. Mayle, J.R. Wilson, and D.N. Schramm, Astrophys. J.
{\bf 322}, 795 (1987); Y.Z. Qian, G.M. Fuller, G.J. Mathews, R.W.
Mayle, J.R. Wilson, S.E. Woosley, Phys. Rev. Lett. {\bf 71}, 1965 (1993).
\bibitem{sawyer}
R.F. Sawyer, Phys. Rev. {\bf D42}, 3908 (1990).
\bibitem{koonin}
A. Schafer and S.E. Koonin, Phys. Lett. B {\bf 185}, 417 (1987).
\bibitem{wick}
W. C. Haxton and W-M. Zhang, Phys.  Rev.  D {\bf 43}, 2484  (1991).
\bibitem{denvar}
P.I. Krastev and A. Yu. Smirnov, Phys. Lett. B {\bf 226}, 341 (1989).
\bibitem{nic}
A. Nicolaidis, Phys. Lett. B {\bf 262}, 303 (1991).
\bibitem{enq}
K. Enqvist and V. Semikoz, Phys.\ Lett.\  B {\bf 312}, 310 (1993).
\bibitem{kam}
N. G. Van Kampen, ``Stochastic Processes in Physics and Chemistry,'' North
Holland, (1990).
\bibitem{bac}
J. N. Bahcall, W. F. Huebner, S. H. Lubow, P. D. Parker, and R. K. Ulrich,
Rev.\ Mod.\ Phys.\ {\bf 54}, 767 (1982), J. Bahcall and R. Ulrich,
Rev.\ Mod.\ Phys.\ {\bf 60}, 297 (1988); J.N. Bahcall and M.H.
Pinsonneault, Rev. Mod. Phys. {\bf 64}, 885 (1992).
\bibitem{sof}
A. B. Balantekin, P. J. Hatchell, F. Loreti, Phys. Rev.\ D {\bf 41}, 3583
(1990).
\bibitem{schram}
X. Shi, D. N. Schramm, R. Rosner, and D. S. Dearborn, Commun. Nucl. Part.
Phys.\ {\bf 21}, 151 (1993).
\bibitem{raf}
G. G. Raffelt, Phys.\ Rev.\ Lett.\ {\bf 64}, 2856 (1990); Phys.\ Rep.\
{\bf 198}, 1 (1990).
\bibitem{me}
H. Minakata and H. Nunokawa, Phys.\ Rev.\ Lett.\ {\bf 63}, 121 (1989);
A. B. Balantekin and F. Loreti, Phys. Rev.\ D {\bf 45},  1059 (1992).
\bibitem{fl}
F.N. Loreti, Y. Qian, A.B. Balantekin, and G. Fuller, in preparation.

\end{thebibliography}
\end{document}